\documentclass[epj]{svjour}
\usepackage{latexsym}
\usepackage{graphics}
\usepackage[latin1]{inputenc}
\usepackage{rotating}
\begin{document}
\renewcommand{\textfraction}{0.00000000001}
\renewcommand{\floatpagefraction}{1.0}
\title{Quasi-free photoproduction of $\eta$-mesons off
the deuteron}
\author{ 
        I.~Jaegle\inst{1},
	B.~Krusche\inst{1},
        A.V.~Anisovich\inst{2,3},	
        J.C.S.~Bacelar\inst{4},
        B.~Bantes\inst{5},
        O.~Bartholomy\inst{2},
        D.E.~Bayadilov\inst{2,3},
        R.~Beck\inst{2},
        Y.A.~Beloglazov\inst{3},
        R.~Castelijns\inst{4},
        V.~Crede\inst{6},
	M.~Dieterle\inst{1},
        H.~Dutz\inst{5},
        D.~Elsner\inst{5},
        R.~Ewald\inst{5},
	F.~Frommberger\inst{5},
        C.~Funke\inst{2},
        R.~Gothe\inst{5,8},
        R.~Gregor\inst{7},
        A.B.~Gridnev\inst{3},
        E.~Gutz\inst{2},
	W.~Hillert\inst{5},
        S.~H\"offgen\inst{5},
        P.~Hoffmeister\inst{2},
        I.~Horn\inst{2},
        J.~Junkersfeld\inst{2},
        H.~Kalinowsky\inst{2},
        S.~Kammer\inst{5},
	I.~Keshelashvili\inst{1},
        V.~Kleber\inst{5},
        Frank Klein\inst{5},
        Friedrich Klein\inst{5},
        E.~Klempt\inst{2},
        M.~Konrad\inst{5},
        M.~Kotulla\inst{1},
        M.~Lang\inst{2},
        H.~L\"ohner\inst{4},
        I.V.~Lopatin\inst{3},
        S.~Lugert\inst{7},
	Y.~Maghrbi\inst{1},
        D.~Menze\inst{5},
	T.~Mertens\inst{1},
        J.G.~Messchendorp\inst{4},
        V.~Metag\inst{7},
        V.A.~Nikonov\inst{2,3},
        M.~Nanova\inst{7},
        D.V.~Novinski\inst{2,3},
        R.~Novotny\inst{7},
	M. Oberle\inst{1},
        M.~Ostrick\inst{5,9},
        L.M.~Pant\inst{7,10},
        H.~van~Pee\inst{2,7},
        M.~Pfeiffer\inst{7},
	F.~Pheron\inst{1},
        A.~Roy\inst{7,11},
        A.V.~Sarantsev\inst{2,3},	
        S.~Schadmand\inst{7,13},
        C.~Schmidt\inst{2},
        H.~Schmieden\inst{5},
        B.~Schoch\inst{5},
        S.V.~Shende\inst{4,12},
        V.~Sokhoyan\inst{2},
        A.~S{\"u}le\inst{5},
        V.V.~Sumachev\inst{3},
        T.~Szczepanek\inst{2},
        U.~Thoma\inst{2,7},
        D.~Trnka\inst{7},
        R.~Varma \inst{7,11},
        D.~Walther\inst{5},
        C.~Wendel\inst{2},
	D.~Werthm\"uller\inst{1},
	\and L.~Witthauer\inst{1}
\newline(The CBELSA/TAPS collaboration)
\mail{B. Krusche, Klingelbergstrasse 82, CH-4056 Basel, Switzerland,
\email{Bernd.Krusche@unibas.ch}}
}
\institute{Departement Physik, Universit\"at Basel, Switzerland
           \and Helmholtz-Institut f\"ur Strahlen- und Kernphysik
                der Universit\"at Bonn, Germany
	   \and Petersburg Nuclear Physics Institute, Gatchina, Russia	
           \and KVI, University of Groningen, The Netherlands
           \and Physikalisches Institut der Universit\"at Bonn, Germany
           \and Department of Physics, Florida State University, Tallahassee,
           USA
           \and II. Physikalisches Institut, Universit\"at Giessen, Germany
	   \and present address: University of South Carolina, USA
           \and present address: University of Mainz, Germany
           \and on leave from Nucl. Phys. Division, BARC, Mumbai, India
           \and on leave from Department of Physics, Indian Institute of Technology
	        Mumbai, India
           \and present address: Department of Physics, University of Pune, Pune, India
	   \and present address: Institut f\"ur Kernphysik, Forschungszentrum J\"ulich, Germany
}
\authorrunning{I. Jaegle et al.}
\titlerunning{Quasi-free photoproduction of $\eta$-mesons off the deuteron}

\abstract{Precise data for quasi-free photoproduction of $\eta$ mesons off the 
deuteron have been measured at the Bonn ELSA accelerator with the combined
Crystal Barrel/TAPS detector for incident photon energies up to 2.5 GeV.
The $\eta$-mesons have been detected in coincidence with recoil protons and
neutrons. Possible nuclear effects like Fermi motion and re-scattering
can be studied via a comparison of the quasi-free reaction off the bound proton
to $\eta$-production off the free proton. No significant effects beyond 
the folding of the free cross section with the momentum distribution of the
bound protons have been found. These Fermi motion effects can be removed 
by an analysis using the invariant mass of the $\eta$-nucleon pairs
reconstructed from the final state four-momenta of the particles. 
The total cross section for quasi-free $\eta$-photoproduction off the
neutron reveals even without correction for Fermi motion a pronounced bump-like
structure around 1~GeV of incident photon energy, which is not observed for 
the proton. This structure is even narrower in the invariant mass spectrum 
of the $\eta$-neutron pairs. Position and width of the peak in the 
invariant mass spectrum are $W\approx 1665$ MeV and FWHM $\Gamma\approx 
25$ MeV. The data are compared to the results of different models.
 % end of PACS codes
} %end of abstract
\maketitle

\section{Introduction}
The excitation spectrum of the nucleon is one of the most important testing
grounds for our understanding of the strong interaction at the few-GeV scale,
where perturbative methods cannot be applied. It plays a similar role for
the strong interaction as atomic spectra do for the electromagnetic interaction.
However, so far, on the theory side, the only direct connection between baryon 
properties and Quantum Chromodynamics (QCD) has been established with the 
numerical methods of lattice gauge theory. The progress in this field was 
tremendous during the last few years for the ground state properties of hadrons 
\cite{Duerr_08}. But only very recently also first results for excited states 
\cite{Bulava_10} going beyond earlier quenched approximations 
\cite{Burch_06,Basak_07} became available. 

In a more indirect way, experimental observations and QCD are connected via
QCD inspired quark models. However, in spite of their phenomenological
successes, the basis of these models is still not well anchored.  
There is neither consent about the effective degrees of freedom nor about 
the residual quark - quark interactions (see e.g. Ref. 
\cite{Capstick_00,Klempt_10} 
for detailed reviews). Apart from the standard, non-relativistic quark 
model with three equivalent valence quarks, also models with a quark-diquark
structure (see e.g. \cite{Anselmino_93}), algebraic approaches 
\cite{Bijker_94}, and flux-tube models \cite{Isgur_85} have been proposed, all 
with different internal degrees of freedom, giving rise to different excitation 
schemes. In a radically different picture of hadrons, it was even attempted to 
generate all excited states by chirally coupled channel dynamics  
\cite{Kolomeitsev_04}, leaving only the ground-state multiplets as genuine 
$qqq$ states. 

So far, comparison of the experimentally known excitation spectrum of the nucleon 
to model predictions does not clearly favor any of the different models,
but reveals severe difficulties for all of them. The ordering of some of the lowest 
lying states is not reproduced. In particular, the N(1440)P$_{11}$ (`Roper') 
resonance and the first
excited $\Delta$, the P$_{33}$(1600), which in the quark model belong to the 
N=2 oscillator shell, appear well below the states from the N=1 shell. 
Furthermore, even the models with the fewest effective degrees of 
freedom predict many more states than have been observed, which is known as the 
`missing resonance' problem.

Since most states have been observed with elastic scattering of charged pions
it is possible that the database is biased towards states that couple 
strongly to $\pi N$. As an alternative, photon induced reactions, which nowadays can
be experimentally investigated with comparable precision as hadron induced 
reactions, have
moved into focus. In order to avoid bias due to the resonance decay properties, 
a large effort has been made during the last few years at tagged photon 
facilities to study many different final states \cite{Burkert_04,Krusche_03}. 
So far, these experiments have mostly concentrated on the free proton. However, 
data from the neutron is also important, because it reveals the iso-spin 
composition of the electromagnetic excitation amplitudes. In extreme cases 
$\gamma NN^{\star}$ couplings may even be completely forbidden due to SU(6) 
selection rules \cite{Moorehouse_66}. Although due to the non-negligible 
spin-orbit mixing in the wave functions they are not strictly forbidden in more 
realistic models, they remain suppressed and can be better studied using neutron
targets. Such experiments are of course complicated by the non-availability of 
free neutrons as targets, requiring coincident detection of recoil neutrons 
from light target nuclei and reaction models taking into account possible 
nuclear effects on the observed cross sections. 

During the last few years the CBELSA/TAPS collaboration has started in Bonn
an extensive program for the study of quasi-free meson production 
reactions off the deuteron, including the $n\eta$ \cite{Jaegle_08}, 
$n\eta '$ \cite{Jaegle_11}, $n\omega$, $n\pi^0\pi^0$, and $n\eta\pi^0$ final states.
These are reactions with only neutrons
and photons in the final state, which can only be investigated with highly
efficient electromagnetic calorimeters covering almost the full solid angle.
In the present paper we discuss the experimental details and analysis procedures 
and summarize the results for the $\eta$-photoproduction. Some results from this
reaction have already been presented in a letter \cite{Jaegle_08}, 
the results for the $\eta '$-channel have been published recently, and
the results from the other channels will be published elsewhere.   

\section{Quasi-free photoproduction of $\eta$-mesons}
  
Photoproduction of $\eta$-mesons off the free proton has been previously
studied in detail
\cite{Krusche_95,Ajaka_98,Bock_98,Armstrong_99,Thompson_01,Renard_02,Dugger_02,Crede_05,Nakabayashi_06,Bartholomy_07,Elsner_07,Denizli_07,Crede_09,Williams_09,Sumihama_09,McNicoll_10}
from the production threshold at $\approx$707 MeV up to incident photon energies
of 2.8 GeV. In the threshold region this reaction is completely dominated by
the photoexcitation of the S$_{11}$(1535) resonance \cite{Krusche_97}. A detailed 
analysis of the angular distributions \cite{Krusche_95} revealed a small contribution 
of the D$_{13}$(1520) via an interference with the leading $E_{0^+}$-multipole of the 
S$_{11}$ excitation. The effect is more pronounced for the beam asymmetry measured 
with linearly polarized photons \cite{Ajaka_98,Elsner_07} and an analysis in the framework
of the `Eta-MAID' model \cite{Chiang_02} allowed the extraction of the tiny 
$N\eta$ branching ratio (0.23$\pm$0.04 \%) \cite{PDG} of the D$_{13}$ resonance.
In the range of the second S$_{11}$ resonance, the S$_{11}$(1650), all models find 
a destructive interference between the two S$_{11}$ states. The situation is less 
clear above this range, where  different analyses like Eta-MAID \cite{Chiang_02} 
and Bonn-Gatchina (BoGa) \cite{Anisovich_05} propose different resonance contributions
(see e.g. \cite{Elsner_07}).

So far, the isospin degree of freedom was almost exclusively explored in the
excitation range of the first S$_{11}$ resonance. The results of quasi-free 
production off the deuteron and $^4$He with and without detection of
recoil nucleons \cite{Krusche_95a,Hoffmann_97,Hejny_99,Weiss_03} are all consistent 
with:
\begin{equation}
\frac{\sigma_n}{\sigma_p}\approx \frac{2}{3}
\label{eq:rat}
\end{equation}   
where $\sigma_n$ and $\sigma_p$ are the cross sections for $\eta$-photoproduction
off neutrons and protons, respectively. The resulting ratio of the helicity 
couplings $A_{1/2}^n$ and
$A_{1/2}^p$ for the S$_{11}$(1535) is \cite{Krusche_03}
\begin{equation}
|A_{1/2}^n|/|A_{1/2}^p|=0.82\pm0.02\;\;\; .
\label{eq:helrat}
\end{equation}
The investigation of coherent $\eta$-photoproduction off the deuteron
\cite{Hoffmann_97,Weiss_01}, $^{3}$He \cite{Pfeiffer_04}, and $^{4}$He
\cite{Hejny_99}, and the comparison of the interference terms in the
angular distributions of quasi-free production off the proton and the neutron
\cite{Weiss_03} have shown that the S$_{11}$(1535) excitation is dominantly
iso-vector, so that \cite{Krusche_03}
\begin{equation}
A_{1/2}^{IS}/A_{1/2}^p=0.09\pm0.01\;\;\; 
\end{equation} 
where $A_{1/2}^{IS}$ is the iso-scalar part of the amplitude.

\begin{figure}[th]
\resizebox{0.49\textwidth}{!}{%
% normal version
  \includegraphics{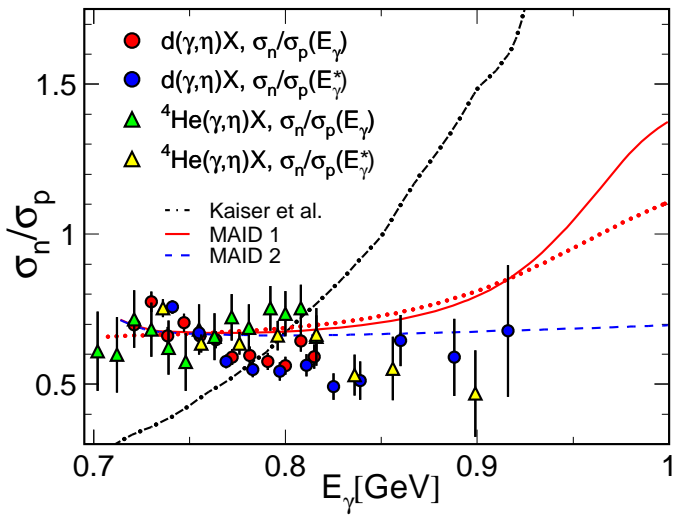}
}
\caption{Comparison of measured $\sigma_n/\sigma_p$ ratios from 
\cite{Hejny_99,Weiss_03} to model predictions. Eta-Maid model \cite{Chiang_02}
(MAID 1 (solid curve): full model, dotted curve: full model after folding with
momentum distributions of bound nucleons, MAID 2 (dashed curve): 
only S$_{11}$(1535) and background). Dash-dotted curve: chirally coupled channel 
dynamics, Kaiser et al. \cite{Kaiser_95,Kaiser_97}. 
$E_{\gamma}^{\star}$: effective photon energy corrected for Fermi 
momentum. 
}
\label{fig:old_ratio}       % Give a unique label
\end{figure}

Previously measured and predicted cross section ratios are summarized 
in Fig. \ref{fig:old_ratio}. The Eta-MAID model \cite{Chiang_02} agrees well
with the 2/3 ratio in the S$_{11}$(1535) range, but predicts for higher 
incident photon energies a significant rise due to the contributions from 
other resonances. This effect should still be visible for quasi-free cross 
sections measured for nucleons bound in the deuteron, which is demonstrated
by the dotted curve in Fig. \ref{fig:old_ratio}. The largest contribution
to the rise comes in this model from the D$_{15}$(1675) resonance, which is one 
of the states which due to the Moorehouse selection rules \cite{Moorehouse_66} 
should have much larger electromagnetic couplings to the neutron than to 
the proton. However, the $N\eta$ branching ratio of this state
was determined in the model from a fit to the proton data as 17 \%. This rather 
large value is in conflict with other results; PDG \cite{PDG} gives only an upper 
limit of $<$ 1\%. 

A strong rise of the ratio to higher incident photon energies was also predicted
in the framework of a chiral coupled channel model by Kaiser and collaborators
\cite{Kaiser_95,Kaiser_97}. But in this case the rise is not related to the
contribution of higher lying resonances, it is related to the properties of the
S$_{11}$(1535), which in this model is not interpreted as a genuine $qqq$ state 
but as a dynamically generated $K\Sigma$-molecule like state. 

Finally, also in the framework of the chiral soliton model \cite{Polyakov_03,Arndt_04} 
a state was predicted in this energy range, which has a much stronger photon 
coupling to the neutron than to the proton {\em and} a large decay branching 
ratio into $N\eta$. This state is the nucleon-like member of the conjectured 
anti-decuplet of pentaquarks, which would be a P$_{11}$ state. 
Exact SU(3)$_{F}$ would forbid the photo-excitation of the proton to the 
proton-like member of the anti-decuplet. But even after accounting
for SU(3)$_{F}$ violation the chiral soliton model predicts \cite{Polyakov_03}
that the photo-excitation of this state is suppressed on the proton and should 
mainly occur on the neutron. Kim et al. \cite{Kim_05} have calculated 
the magnetic transitions moments for the anti-decuplet states and found a
considerable enhancement for the excitation of the nucleon-like state on the
neutron with respect to the proton. 

Strong efforts have recently been undertaken
at various facilities (GRAAL in Grenoble \cite{Kuznetsov_07}, ELSA in Bonn
\cite{Jaegle_08} and LNS in Sendai \cite{Miyahara_07}) to extract reliable 
results for the $\gamma n\rightarrow n\eta$ reaction at higher incident photon 
energies. The somewhat unexpected finding in all experiments is not only a 
significant rise of the cross section on the neutron with rising 
incident photon energy, but a fairly narrow peak in it, which has no 
counterpart in the reaction off the proton. 

The nature of this structure is still
unknown and many different suggestions have been put forward in the literature.
They include various types of coupled channel effects involving known nucleon
resonances \cite{Shklyar_07,Shyam_08} or the opening of production thresholds
\cite{Doering_10}, but also scenarios with contributions from intrinsically
narrow states have been discussed \cite{Fix_07,Anisovich_09}. Very recently,
Kuznetsov et al. \cite{Kuznetsov_11} have reported a narrow structure in
Compton scattering off the quasi-free neutron with a similar mass and width, although
not with large statistical significance ($\approx$4.6$\sigma$).
Such an observation would make explanations with complicated interference and
threshold effects less likely since there is no good reason why they should appear
similar for such different reactions. A nucleon resonance with strong 
photon coupling to the neutron, on the other hand, naturally would be expected 
to show up also in Compton scattering off the neutron. 

In this paper, we will summarize and discuss in detail the results from the
ELSA experiment reported partly in \cite{Jaegle_08} and in addition present a 
new analysis which removes the effects of Fermi motion from the data through a
kinematic reconstruction of the involved nucleon momenta. In this way, more
stringent constraints can be put on the intrinsic width of the peak-like structure.

The paper is organized in the following way. The experimental setup is described  
in Sec. \ref{setup}. Details of the analysis are discussed in Sec. \ref{analysis}.
This includes the calibration of all detector components, the identification of
photons, reconstruction of mesons, and detection of recoil nucleons, the absolute
normalization of the cross sections and a thorough discussion of systematic uncertainties.
The results are summarized in Sec. \ref{results}. We first compare the data to previous
results from quasi-free photoproduction off the deuteron and the results 
for the quasi-free proton to free proton data. In the following two subsections
the results for quasi-free photoproduction are first discussed in dependence of the
incident photon energy. In this analysis the width of narrow structures is dominated 
by nuclear Fermi motion. In a second analysis results are constructed in dependence 
of the nucleon - meson invariant mass in the final state calculated from the
four vectors of the observed particles. In this way, Fermi motion effects are removed 
and only instrumental resolution must be considered.

\begin{figure*}[th]
\resizebox{1.0\textwidth}{!}{%
% normal version
  \includegraphics{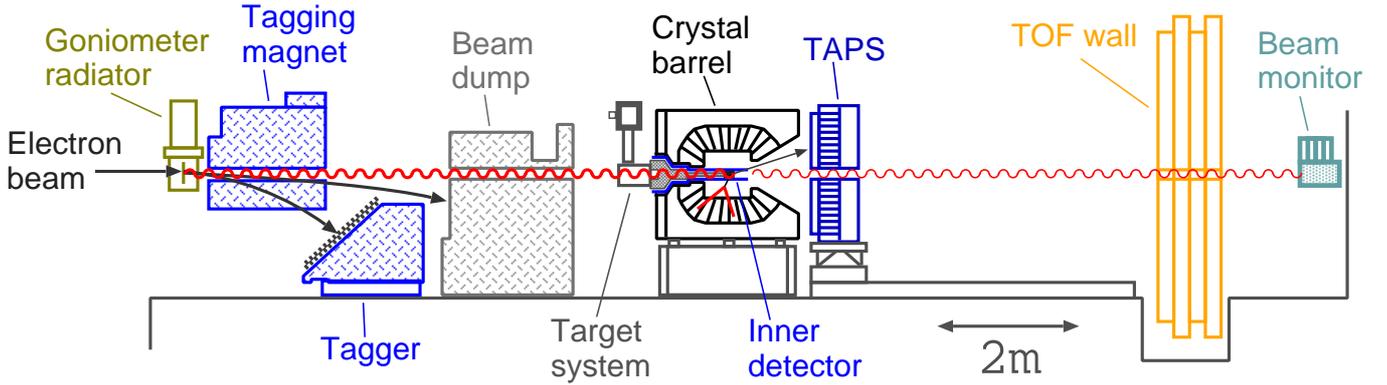}
}
\caption{Overview of the experimental setup. The electron beam enters from the left side.
Scattered electrons were detected in the counters of the tagging spectrometer (cf. Fig.
\ref{fig:tagger}). The liquid deuterium target was mounted in the center of the Crystal
Barrel. The forward range was covered by the TAPS detector. The Time-of-flight wall
was mounted, but not used in the experiment. Beam intensity was monitored at the end of the
beam line.} 
\label{fig:setup}       % Give a unique label
\end{figure*}

\section{Experimental setup}
\label{setup}

The experiment was performed at the electron stretcher accelerator facility
ELSA in Bonn \cite{Husmann_88,Hillert_06}, which can deliver electron beams with 
intensities of a few nA for energies up to 3.5 GeV. The overall setup of the 
experiment is shown in Fig. \ref{fig:setup}. The incoming electron beam is
impinging on the radiator mounted in a goniometer. For the measurement discussed
here, electron beam energies of 2.6~GeV and 3.2~GeV were used. The largest part 
of the 3.2~GeV beam time was done with a copper radiator foil of 0.3~\% radiation 
lengths thickness, producing unpolarized bremsstrahlung. For the rest of this 
beam time and for the beam times with 2.6 GeV electron energy a diamond radiator 
was used to produce coherent bremsstrahlung with maximum linear polarization 
around 1 GeV incident photon energy (see \cite{Elsner_09} for details about 
linear polarization of the photon beam). 

\begin{figure}[th]
\resizebox{0.5\textwidth}{!}{%
% normal version
  \includegraphics{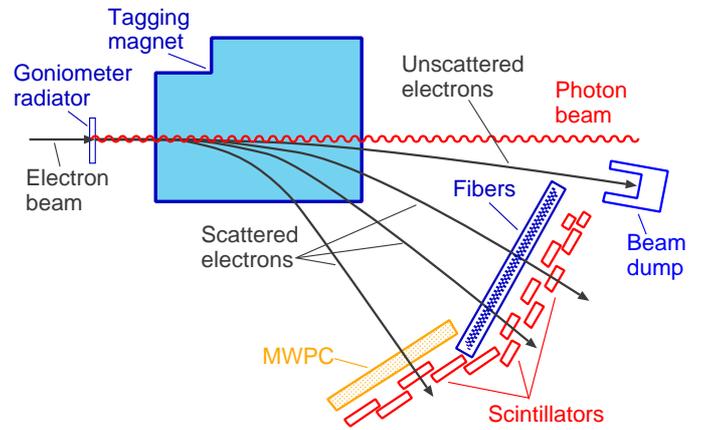}
}
\caption{Setup of the tagging spectrometer. Unscattered electrons stop in the beam-dump.
Scattered electrons pass a two layer detection system. Scintillating fiber detectors
(good position resolution) and scintillator bars (good time resolution) cover 
photon energies up to 80 \% of the electron beam energy. The part covered by a 
multiple wire chamber (MWPC) (low energy electrons corresponding to
photon energies above 80 \% of the electron beam energy) was not used in the experiment.  
}
\label{fig:tagger}       % Give a unique label
\end{figure}

Electrons which have emitted bremsstrahlung, are then deflected downwards by a
dipole magnetic field onto the focal plane of the tagging system, where energy and 
timing information is extracted. Unscattered electrons are stopped in the beam dump.
The bremsstrahlung photons are almost co-linear with the incident electron beam, 
pass through a hole in the magnet yoke and
irradiate the liquid deuterium target mounted in the center of the Crystal 
Barrel detector. The core of the detector system of the tagging spectrometer 
(see Fig. \ref{fig:tagger}) are 14 partially overlapping plastic scintillator bars 
each with photo-multipliers at both ends. This system covers 22~\% to 92~\% of the 
incoming electron beam energy $E_{o}$. A second layer has been added in front of 
the bars for better energy resolution. A scintillating fiber detector 
(2 $\times$ 240 fibers arranged in two layers to partly overlap) covers photon
energies from 18~\% to 80 \% of $E_{o}$. This defined the maximum tagged photon energy 
of 2.5~GeV.
The fiber detector provides an energy bin width of $\approx$1.5~\% for the lowest 
incident photon energies and $\approx$0.1~\% at the high energy. In principle,
the range from  80 \% to 92 \% of $E_o$ can be tagged with an additional wire chamber.
The chamber was, however, not used in the present experiment since due to the small
cross section the statistical quality of data in this range was not sufficient.
 
The target consisted of a vertically mounted cryostat attached to a tube entering
the Crystal Barrel detector from the upstream side. The target cell itself was a
capton cylinder (0.125 mm foil thickness) with a diameter of 3~cm and length of
5.275~cm, filled with liquid deuterium (surface thickness 0.26 nuclei/barn).

\begin{figure}[thb]
\resizebox{0.45\textwidth}{!}{%
% normal version
  \includegraphics{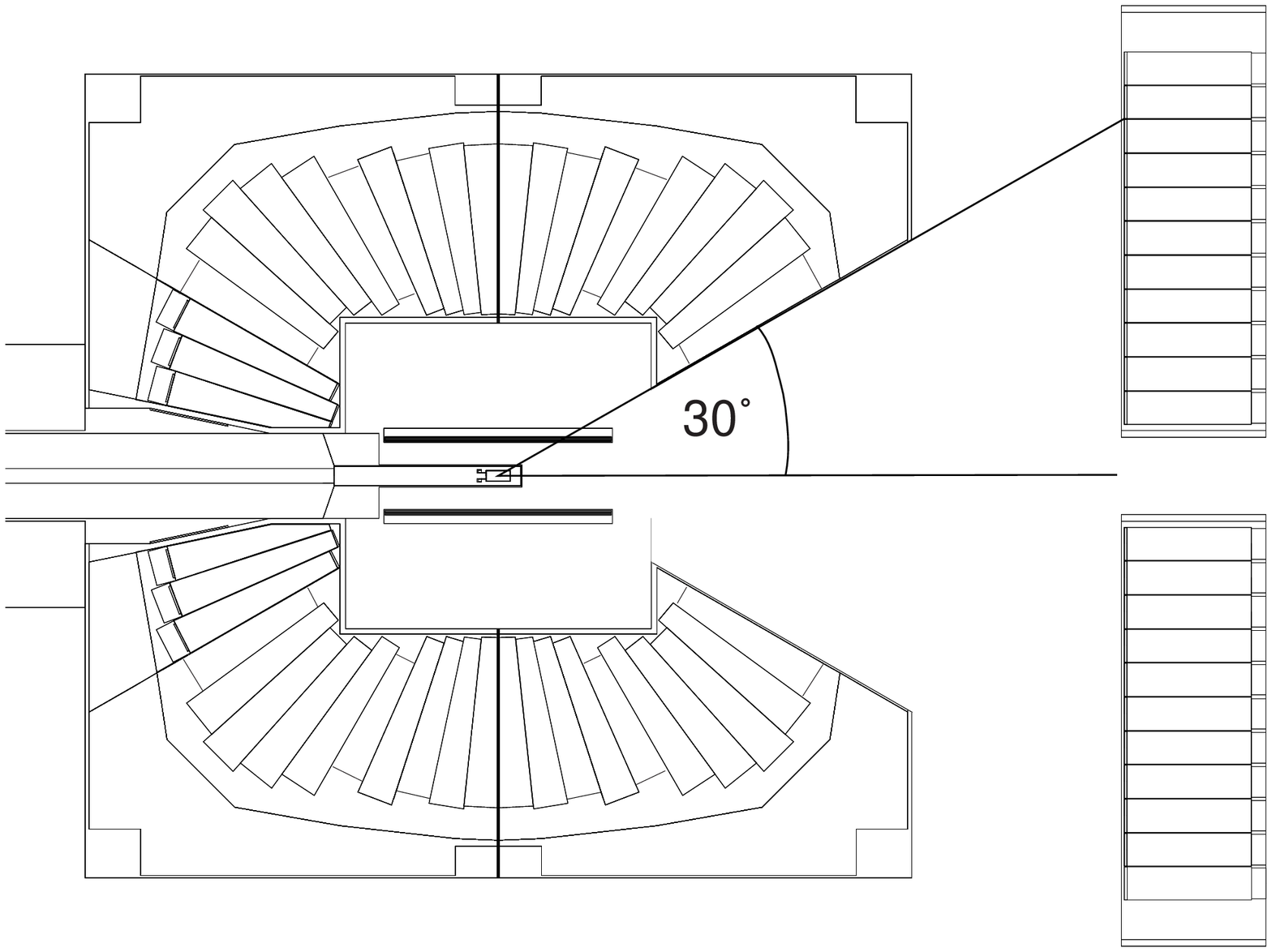}
}
\end{figure}
\begin{figure}[hbt]
\vspace*{0.cm}
\resizebox{0.24\textwidth}{!}{%
% normal version
  \includegraphics{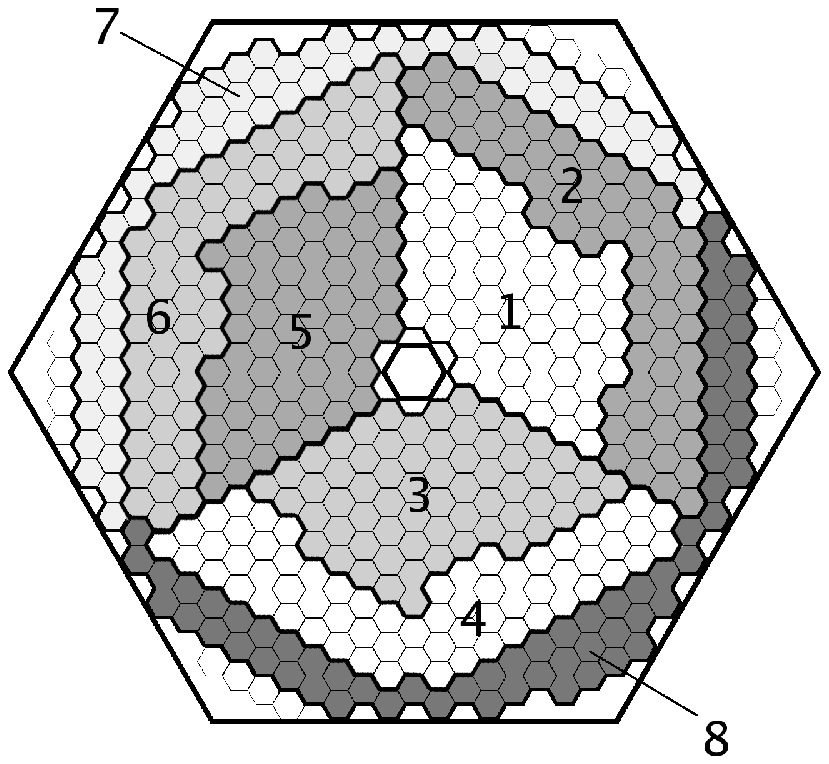} 
}
\resizebox{0.24\textwidth}{!}{%
  \includegraphics{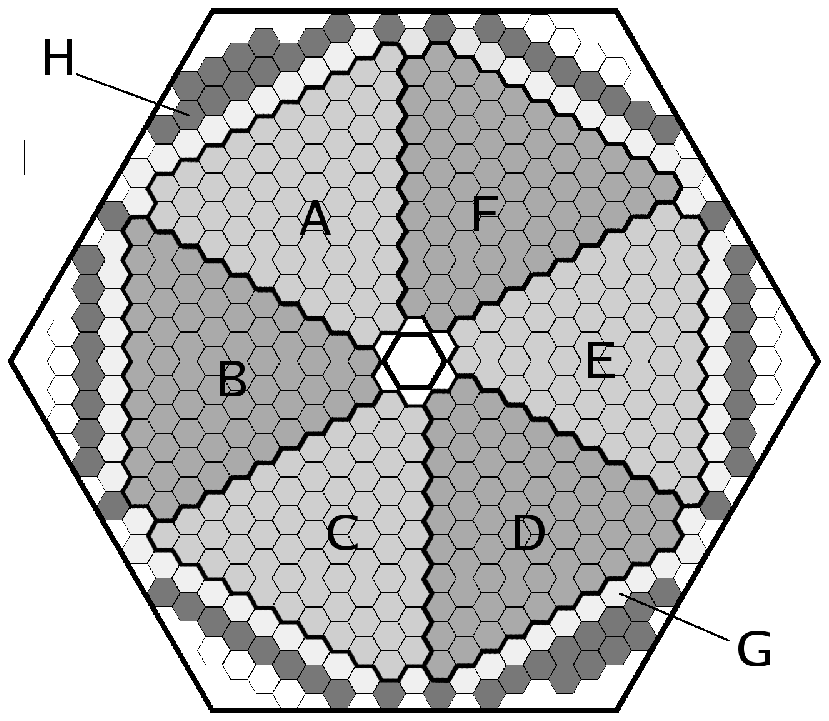}
}
\caption{Arrangement of the Crystal Barrel and TAPS detectors. Upper part:
side view, lower part: front view of the TAPS wall: left hand side: logical
segmentation for the LED-low trigger, right hand side: logical segmentation
for the LED-high trigger (see text).
}
\label{fig:calo}       % Give a unique label
\end{figure}

Reaction products emerging from the target have been detected with a combined
setup (see Fig.~\ref{fig:calo}) of the Crystal Barrel detector \cite{Aker_92} and 
the TAPS detector \cite{Novotny_91,Gabler_94}. In the configuration used, the 
Crystal Barrel consisted of 1290 CsI (Tl) crystals of 16 radiation lengths $X_o$ 
all mounted in a target pointing geometry. It covered the full azimuthal angle for 
polar angles between 30$^{\circ}$ and 168$^{\circ}$. The forward angular range 
was covered by the TAPS detector \cite{Novotny_91,Gabler_94}. This component 
was made of 528 BaF$_2$ crystals of hexagonal shape with an inner diameter of 
5.9 cm and a length of 25 cm corresponding to 12 $X_{o}$. They were arranged 
in a wall-like structure as shown in the lower part of Fig.~\ref{fig:calo}, 
covering polar angles down to 4.5$^{\circ}$. The front face of the BaF$_2$ wall 
was located 1.18 m from the center of the target. 

The two calorimeters have a comparable energy resolution \cite{Aker_92,Gabler_94}
\begin{equation}
\frac{\sigma_E}{E}\approx \frac{2 - 3 \%}{\sqrt[4]{E/GeV}}\;\;\;.
\label{eq:eresol}
\end{equation}
Since the impact points of photons are determined from the center of gravity of the
electromagnetic showers, the angular resolution is better than the granularity of the 
crystals. It is 1.5$^{\circ}$ ($\sigma$) for the CB \cite{Aker_92} for photons 
with energies above 50 MeV and 1.25$^{\circ}$ in TAPS. Angular resolution for recoil
nucleons, which do not produce extended clusters, is closer to the granularity.

Both parts of the calorimeter
were equipped with detectors for the identification of charged particles.
A three-layer scintillating fiber detector (`Inner'-detector) \cite{Suft_05} 
was mounted inside the Crystal Barrel around the target, covering polar angles 
between 28$^{\circ}$ and 172$^{\circ}$. The outer layer (diameter 12.8 cm, 
191 fibers) runs parallel to the z-axis. The middle layer (\O = 12.2 cm, 
165 fibers) is wound anti-clockwise at an angle of 25$^{\circ}$ with respect 
to the z-axis and the inner layer (\O = 11.6 cm, 157 fibers) is wound clockwise
at 25$^{\circ}$. All fibers have 2 mm diameter. This orientation allows the
reconstruction of the spatial coordinates of the intersection point of the
charged particle trajectory with the detector (see \cite{Pee_07} for details).  
The TAPS detector was complemented with a charged-particle-veto (CPV) detector
built of 5 mm thick plastic scintillators of hexagonal shape and the same
dimensions as the front-face of the BaF$_2$ crystals, so that each detector
module had its individual veto detector. The CPV was read out by wave-length
shifting fibers connected to multi-anode photo-multipliers.   

The fast BaF$_2$ modules of the TAPS detector were read out by photo-multipliers,
but the CsI crystals of the Barrel were read by photo-diodes, which do not provide
any time information. This had important consequences for the hardware trigger. 
Signals for the first-level trigger could only be derived from the TAPS 
detector. For this purpose each module was equipped with two
independent leading-edge discriminators, combined in two different ways into
logical groups (see Fig. \ref{fig:calo}). 
For most of them (ring 12 - 5 from outer 
edge to center) a lower threshold was set to $\approx$55 MeV (LED-low). 
It was set to 80 MeV, 135 MeV, 270 MeV for rings 4, 3, 2, respectively. 
The inner-most ring was not used in the trigger.
The LED-high thresholds were set to 70 MeV for rings 9 - 7, rising from
105 MeV (ring 6) to 180 MeV (ring 2). Again, the inner-most ring was not
used in the trigger and the three outer rings (block G) had no leading edge
discriminators for the high threshold.
The first-level trigger then included 
two components: at least two LED-low discriminators from different logical 
sectors above threshold, or at least one LED-high discriminator above threshold.
In the second case, a second-level trigger from the FAst Cluster Encoder 
(FACE) of the Crystal Barrel, indicating at least two separated hits in the 
Barrel, was required in addition. All first level triggers thus required
detection of one or two photons in TAPS, which covered only a small part of the
solid angle. Therefore, only reactions with relatively high photon multiplicity
like $\gamma d\rightarrow np\pi^0\pi^0\rightarrow np 4\gamma$,
$\gamma d\rightarrow np\eta\rightarrow np 3\pi^0\rightarrow np 6\gamma$, or
$\gamma d\rightarrow np\eta '\rightarrow np\eta 2\pi^0\rightarrow np 6\gamma$
could be recorded. Two-photon decays of single meson production reactions
like $\gamma d\rightarrow np\pi^0\rightarrow np 2\gamma$ or
$\gamma d\rightarrow np\eta\rightarrow np 2\gamma$ were not taken. 
It should be mentioned that this restriction of the detector setup
applies only to reactions off the (quasi)-free neutron. In measurements off
the free proton, the recoil protons provide trigger signals from the Inner-detector 
(or TAPS). The TAPS modules were additionally equipped with constant fraction
discriminators (CFD), with thresholds around 10 MeV, which were used
for the high resolution measurement of the time and generated the read-out 
pattern of the  detector matrix. 

The last component of the setup is the so-called $\gamma$-veto detector
at the end of the photon-beam line, which consisted of nine lead-glass crystals,
and counted the photons which passed through the target without interaction.
It was used to monitor the time dependence of the photon-beam intensity.
Furthermore, it was used to measure the tagging efficiency, i.e. the
fraction of photons, correlated with electrons, which impinged on the target. 
This was done in special tagging efficiency 
runs, where at reduced beam intensity the tagger was used as trigger and the number 
of scattered electrons counted in the tagger was compared to the number of
coincident photons impinging on the $\gamma$-veto. 
   
\section{Data analysis}
\label{analysis}

In this chapter we discuss the different analysis steps of the experiment,
from the most basic calibration procedures for the different detector 
components, over the identification of photons and particles, the 
identification of specific reaction channels, to the absolute normalization 
of cross sections. Although in this paper only final states with $\eta$ mesons
are discussed, a large part of this chapter applies also to the other,
simultaneously measured reactions, which will be reported elsewhere. Therefore
we discuss in this paper the analysis steps in some detail also in view of 
possible systematic uncertainties. A full account of all analysis procedures is 
given in \cite{Jaegle_07}.

\subsection{Calibration procedures}

\subsubsection{Tagging system}

The tagging system has two tasks: the event-by-event definition of the
incident photon energy via a coincidence between the focal plane detectors 
and the reaction detector and the monitoring of the photon flux
from the counting of the deflected electrons. 

For the present analysis energy and timing information was obtained with the 
scintillating fiber detector. Since for any scattered electron the bending
radius $\rho$ in the magnetic field $B$ is related to its momentum $p$ by
\begin{equation}
\rho = \frac{p}{Bq}
\end{equation}
where $q$ is the electron charge, its energy follows from the position in the 
focal plane detector, i.e. the number of the responding fiber. The energy calibration
of the scintillating fiber detector was done in two steps. As a starting
point a polynomial calibration function was calculated from the measured field
map of the dipole magnet and the positions of the scintillating fibers.
Initially, this calibration was done for an electron beam energy of
3.2 GeV. For other beam energies, the magnetic field was adjusted such that
the unscattered electrons always followed the same trajectory so that the
calibration function could be simply scaled by the electron beam energy. 
This calibration was checked by direct injection of a very low intensity 
electron beam with removed radiator. At a fixed magnetic field setting for 
3.2 GeV ($B$=1.413 T) ELSA beams with four different energies 
(1.2, 1.5, 2.0, 2.5 GeV) were used.

\begin{figure}[th]
\resizebox{0.47\textwidth}{!}{%
% normal version
  \includegraphics{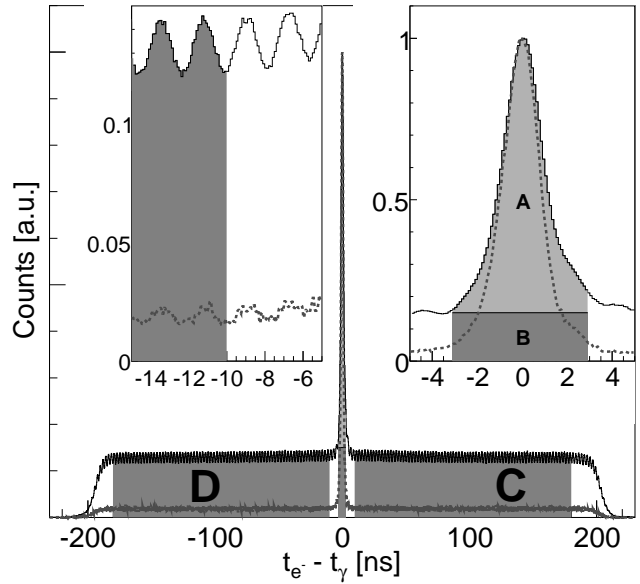}
}
\caption{Relative TAPS-tagger time spectrum for all events with photons in 
TAPS (solid histograms) and for events with $\eta$-decay photons in TAPS
(dashed histograms).
The 2 ns time structure of the beam is reflected in the fine structure of the
accidentals. 
The shaded areas indicate the prompt (A) and random (B, C, D) coincidence 
areas (see text).
}
\label{fig:time}       % Give a unique label
\end{figure}

A further cross check was done with the position of the peaks from coherent
bremsstrahlung produced in a diamond crystal, which was used for measurements
with linearly polarized photon beams. The systematic uncertainty of this
calibration is on the order of the energy bin widths of the fiber detector.  

The measurement of the relative timing between the scattered electrons
and the reaction products was done with the focal plane scintillating fiber
detector and the TAPS detector. The modules of both detector systems are
equipped with individual TDC's. In case of the focal plane detector the start
signal came from the individual fibers and the stop from the trigger signal. 
The TAPS TDC's were started with the trigger signal and stopped 
by the individual CFD signals. The time calibration of the tagger TDC's was 
64 ps/channel. After alignment of all channels a time resolution of 1.6 ns
(FWHM) was achieved for the sum spectrum, which is shown in Fig.~\ref{fig:time}.
The average multiplicity of hits in the focal plane detector per event in the
coincidence time window of 460 ns was $\approx$20, resulting in a significant
random background. After a cut on the prompt peak ($\pm$3 ns), the remaining 
background corresponding to region B in Fig.~\ref{fig:time} was removed from 
all shown spectra in the usual way by subtraction of the events from the 
areas D and C normalized by the ratio of the areas B/(D+C). The random 
background was much less important for events with photons from an identified 
$\eta$-meson (dashed histograms in Fig.~\ref{fig:time}).
Also the small asymmetric tail of the prompt peak (cf. Fig. \ref{fig:time}),
due to particles misidentified as photons, vanishes in this case.

\subsubsection{TAPS}

A precise calibration of the time measurement with TAPS was not only important
for the suppression of random background. It was also the basis of the
measurement of the kinetic energy of recoil nucleons with the time-of-flight
method. The slightly varying gaines of the TDC's were measured for each 
channel with pulser signals of known delay, fed into the electronic chain. 
The alignment of all channels was done with events with two or more photons 
detected in TAPS. For this procedure, photons were identified with the help 
of the veto detector and an invariant mass analysis, which accepted only 
decay photons of $\pi^o$ mesons. The resulting TAPS - TAPS coincidence time 
spectrum had a resolution of 650 ps (much better than the TAPS - tagger timing)
and a peak-to-background ratio of 100. 

The energy calibration was done in three steps. In the first step the energy
deposition of minimum ionizing cosmic myons ($\approx$38.5 MeV) were used for 
a relative calibration of all detector modules. Such a calibration does not
take into account differences in the detector response to the energy deposition
of myons and electromagnetic showers as well as shower leakage. Therefore, in 
the second step the linear term of the calibration was adjusted for all modules
in an iterative procedure with the invariant mass peak of $\pi^0$ mesons.
In the third step, a quadratic correction of the energy calibration was 
introduced using the position of the invariant mass peak of the $\eta$ meson.
A final accuracy of $\pm$1 \% for the position of the $\pi^o$- and $\eta$-peaks
was achieved. More details can be found in \cite{Jaegle_07}.

\subsubsection{Crystal Barrel}

Due to the read-out with photo-diodes, the CB delivers only energy
information. A calibration with cosmic myons is not practical, due to the
different geometries and orientations of the individual modules.
The digitization of the detector signals is done for two different dynamic
ranges. The calibration for the low-gain chain was again done with an
iterative procedure using the $\pi^0$ invariant mass peak. The high gain 
branch was calibrated by the injection of laser light of known intensity
into the crystals. 

\subsection{Identification of particles and reaction channels}

Electromagnetic showers, depending on their energy, will in general
produce signals in an extended `cluster' of scintillator modules. In a first
step of the analysis all hits in the calorimeters were grouped into
`clusters' of adjacent crystals and the energy sums and geometrical centers 
of gravity of the `clusters' were extracted (see \cite{Pee_07} for details). 
In the next step, the TAPS-veto detector and the Inner-detector 
were used to separate neutral hits (photon and neutron candidates) from 
charged hits (proton candidates) in the detectors. In TAPS a hit was assigned 
to `charged' if the veto of any module from the cluster or the veto of any 
neighbor of the central module of the cluster had fired (even if the neighbor 
module itself had no signal above threshold). The latter condition is 
important for charged particles with relatively large impact angles which may 
traverse the edge of a veto but deposit their energy in the neighbor module. 
All other hits were assigned to `neutral'.
In the Barrel, a hit was assigned to `charged', if at least two layers of the
Inner-detector had recorded a hit within $\pm 10^{\circ}$ of azimuthal angle.
The efficiency of the Inner-detector for this condition is 98.4~\% \cite{Suft_05}.
It was assigned to `neutral' if no layer had fired within this azimuthal angle,
which results in a probability of $\approx$ 0.04 \% to misidentify a charged
particle as neutral hit.  Events with hits with one responding layer of the 
Inner-detector were discarded.
 
After these assignments three partly overlapping classes of events were included 
into the analysis. Class (a) included events with exactly six neutral hits and
exactly one charged hit. Class (b) included events with exactly seven neutral 
hits and no charged hit (in this case was additionally required that the 
Inner detector had not fired at all). Class (c) included all events from 
classes (a) and (b) and in addition all events with exactly six neutral hits
and no charged hit.  
Class (a) corresponds to quasi-free production off the proton 
$\gamma d\rightarrow (n)p\eta$ with coincident detection of the 
recoil proton. Class (b) corresponds to quasi-free production off the 
neutron $\gamma d\rightarrow (p)n\eta$ with coincident detection of the recoil 
neutron. Class (c) corresponds to the inclusive reaction 
$\gamma d\rightarrow (np)\eta$, without any condition for the recoil nucleon 
(may be detected but is not required).

\begin{figure}[th]
\resizebox{0.50\textwidth}{!}{%
% normal version
  \includegraphics{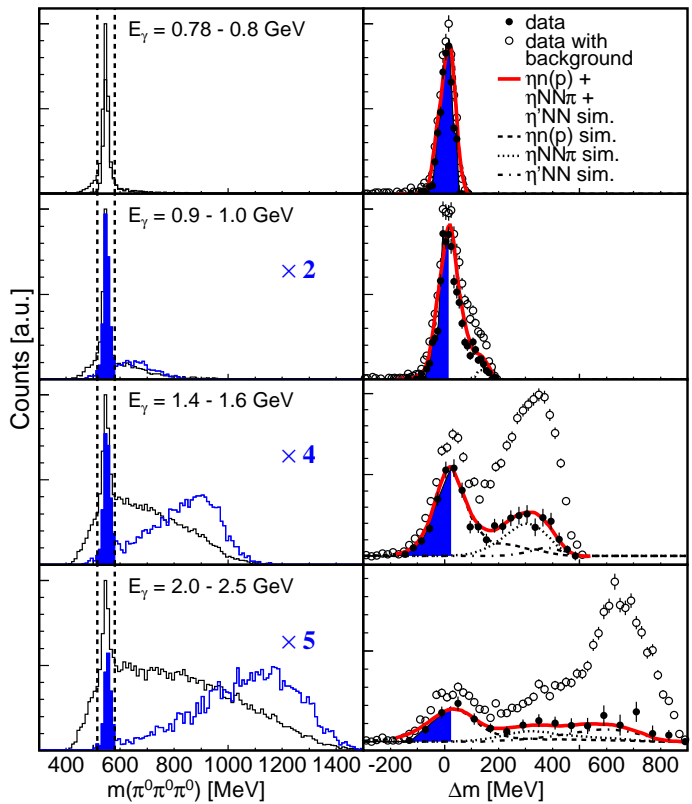}
}
\caption{Invariant (left hand column) and missing mass (right hand column) spectra
for $\eta$-mesons in coincidence with recoil neutrons for three ranges of
incident photon energy. Invariant masses: shaded (blue) signal after missing 
mass cut (scaled up by indicated factors).
Dashed lines: applied invariant mass cut. Missing mass data: open symbols
represents data for indicated cut on invariant mass. Black dots: background subtracted
by fitting invariant mass spectra for each bin of missing mass.
Simulations: dashed (dotted, dash-dotted) curves: simulation of $\eta$ 
($\eta\pi$, $\eta '$) final states. Solid (red) curves: sum of simulations. 
Shaded (blue) areas: accepted events.    
}
\label{fig:mima}       % Give a unique label
\end{figure}

These events were subjected to a combined
invariant and missing mass analysis. In the first step of the invariant mass
analysis the invariant masses of all combinations of three disjunct pairs of
neutral hits were calculated. In the case of six neutral hits (events with
proton candidate or without candidate for recoil nucleon) these are 15 different 
combinations among which the `best' combination was chosen by a $\chi^2$-test,
minimizing
\begin{equation}
\chi^2 = \sum_{k=1}^{3}\frac{(m_{k}(\gamma\gamma)-m_{\pi^0})^2}{(\Delta
m_{k}(\gamma\gamma))^2}
\end{equation}
for all disjunct combinations where $m_{\pi^0}$ is the  pion mass and the 
$m_{\gamma\gamma}$ are the invariant masses of the photon pairs with their 
uncertainties $\Delta m_{\gamma\gamma}$ calculated for each photon pair from 
the energy and angular resolution of the detector.  
For events with seven neutral hits (events with neutron candidate) one must in 
addition loop over the un-paired hit giving rise to 105 combinations.
Once the `best' combination was determined, in all cases a cut between 
110 MeV - 160 MeV was applied to the invariant masses. Only events where all
three pairs of the best combination passed this cut were kept. Subsequently,
for events with seven neutral hits, the residual hit was taken as neutron 
candidate.

The nominal mass of the pion was then used to improve the experimental 
resolution. Since the angular resolution of the detector is much better than
the energy resolution this was done by re-calculating the photon
energies and momenta from the approximation:
\begin{equation}
E'_{1,2}=E_{1,2}\frac{m_{\pi^0}}{m_{\gamma\gamma}}
\end{equation}
where $E_{1,2}$ are the  measured photon energies, $E'_{1,2}$ the re-calculated
energies, $m_{\pi^0}$ is the nominal $\pi^0$ mass, and $m_{\gamma\gamma}$ the
measured invariant mass. The slow $E^{-1/4}$ energy dependence of $\sigma_E/E$
from Eq. \ref{eq:eresol} has been neglected here.  

In the last step, the invariant mass of the six hits assigned as
photons was calculated. The result is shown in Fig. \ref{fig:mima} (left hand side)
for three different ranges of incident photon energy. The figure shows the most
difficult case for events with neutron candidates, where combinatorial
background in the $\chi^2$ analysis of the best combination is higher than
for events with six neutrals only.

After the invariant mass analysis all events were subjected to a missing mass 
analysis, where the recoil nucleon, no matter if a candidate was detected or 
not, was treated as missing particle.
The missing mass $\Delta m$ of the reaction was calculated for quasi-free production
of $\eta$ mesons off nucleons via:
\begin{equation}
\Delta m = \left|{\mbox{\bf P}_{\gamma}+\mbox{\bf P}_{N}-\mbox{\bf P}_{\eta}}\right|
-m_N\ ,
\end{equation}
where $m_N$ is the nucleon mass, ${\mbox{\bf P}_{\gamma}}$, ${\mbox{\bf P}_{N}}$, ${\mbox{\bf P}_{\eta}}$
are the four-momenta of the incident photon, the initial state nucleon 
(assumed to be at rest), 
and the produced $\eta$-meson. The resulting distributions peak around zero. 
They are somewhat broadened by the momentum distribution of the bound
nucleons, which was neglected. Typical results are shown on the right hand side 
of Fig. \ref{fig:mima}. The open symbols correspond to the data after the cuts
on invariant mass shown on the left hand side of the figure have been applied.
At higher incident photon energies background in particular from $\eta\pi$ final
states is visible, where the pion has escaped detection. A smaller background
at the highest incident photon energies is due to the 
$\eta '\rightarrow \eta\pi^0\pi^0$ decay. Cleaner results are obtained when 
the invariant mass spectra are generated for each bin of missing mass 
and fitted with the $\eta$ line shape and a background polynomial. This analysis 
demonstrates (compare open and filled symbols in Fig. \ref{fig:mima}, right hand
side) that the background underneath the $\eta$ invariant mass peak 
contributes mainly to large missing masses. Only a small contribution from
triple $\pi^0$ production appears in the missing mass peak region.
The shape of the missing mass 
structures for the $\eta$, $\eta\pi$, and $\eta '$ final states has been 
generated with a Monte Carlo simulation (see below). The sum of these 
contributions reproduces the measured spectra.

\begin{figure}[ht]
\centerline{
\resizebox{0.50\textwidth}{!}{%
% normal version
 \includegraphics{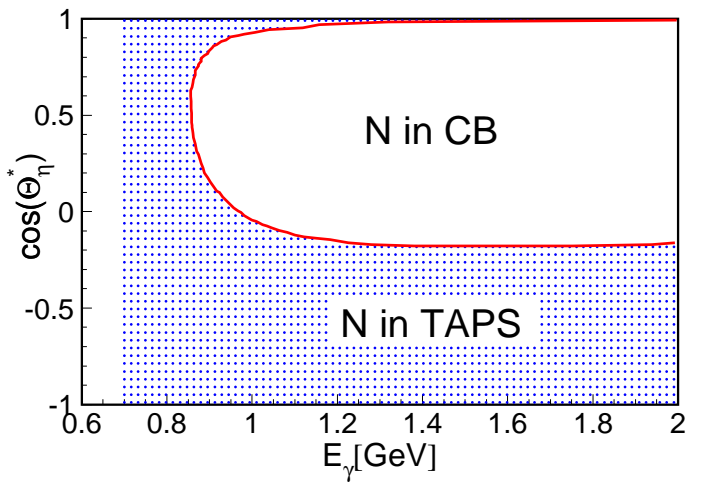}
}}
\caption{Simulated distribution of recoil nucleons depending on incident photon
energy and $\eta$ cm polar angle. Nucleons corresponding to the (blue) shaded area 
are emitted into the solid angle covered by TAPS, nucleons from the white area into
the Barrel. Due to Fermi smearing the borders are not sharp.
}
\label{fig:regions}       % Give a unique label
\end{figure}

The final analysis was done in the following way. For incident photon energies below
800 MeV, where the reaction $\gamma d\rightarrow \eta\pi X$ is kinematically forbidden
and the missing mass spectra are background free, the signals were obtained by
fitting the invariant mass spectra with peak shape and background polynomial.
At higher incident photon energies, where background is no longer negligible, first 
a cut on missing mass was applied, which only accepted events on the left hand side 
of the peak (blue shaded areas in Fig. \ref{fig:mima}, right hand side). 
The corresponding invariant mass spectra are shown 
on the left hand side of the figure as (blue) shaded histograms. The small residual 
background underneath these invariant mass peaks was fitted and removed. This analysis 
was done for each bin of incident photon energy and $\eta$ center-of-momentum (cm) 
polar angle ($\Theta_{\eta}^{\star}$). The missing mass cut was chosen very restrictive, at the price of the loss of 
half the counting statistics. This was done to exclude any background contamination,
which could create artificial structures in the excitation function around 1 GeV 
where the background in the missing mass spectra starts to appear.

Finally, we discuss the assignment of recoil nucleons to the events. Fig.
\ref{fig:regions} shows which recoil nucleons need to be detected in TAPS and 
which in the Barrel, depending on the reaction kinematics. For incident photon 
energies below $\approx$ 850 MeV all recoil nucleons are emitted into the solid
angle covered by TAPS. At higher energies only nucleons corresponding to backward
emission of the $\eta$-mesons (approximately cos($\Theta_{\eta}^{\star}$)$<$-0.2) 
are detected in TAPS.   

\begin{figure}[ht]
\centerline{
\resizebox{0.50\textwidth}{!}{%
% normal version
  \includegraphics{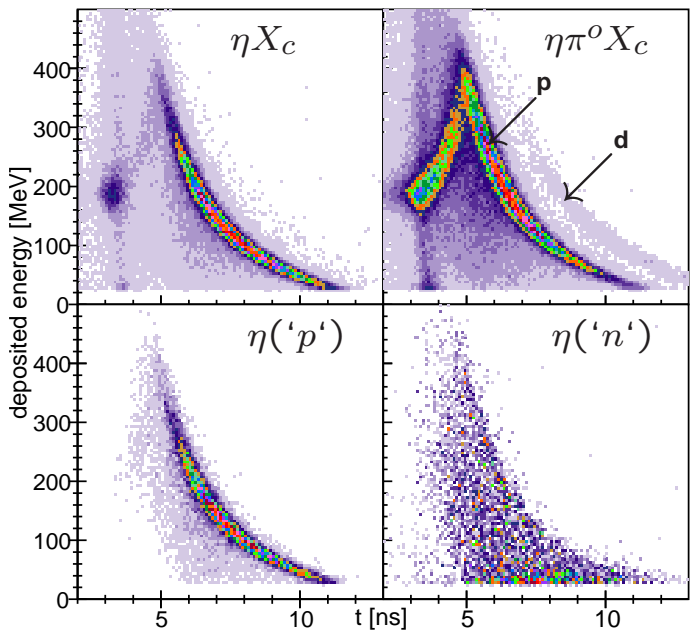}
}}
\caption{Time-of-flight versus energy spectra measured with TAPS for different
event types. Upper left: $\eta\rightarrow 6\gamma$ plus one charged particle,
upper right: $\eta\rightarrow 2\gamma$ and $\pi^o\rightarrow 2\gamma$ plus one 
charged particle, lower left: $\eta\rightarrow 6\gamma$ plus one charged particle
$('p')$ and missing mass cut, lower right: $\eta\rightarrow 6\gamma$ plus one neutral 
particle $('n')$ and missing mass cut.   
}
\label{fig:tof}       % Give a unique label
\end{figure}

Nucleons detected in the Barrel are identified as protons or neutrons as discussed
above with the help of the Inner-detector. Due to the three-layer structure of this
device the probabilities to misidentify a proton as neutron or vice versa are
negligible. Charged pions might be also misidentified as protons. However, such 
events are reliably removed by the missing mass cut since they originate from 
$\eta\pi$ final states.

In case of the TAPS detector, which was placed 1.18~m downstream from the target, an 
independent check of the recoil nucleon identification was done via a 
time-of-flight versus energy analysis of the recoil particles. This is illustrated 
in Fig. \ref{fig:tof}. In the spectrum of charged recoil particles detected in
coincidence with an $\eta$-meson identified by invariant mass (upper left corner) 
a clear proton band is visible. After the missing mass cut (lower left corner)
almost all background is removed. The spectrum of neutron candidates after
application of $\eta$ invariant and missing mass cut (lower right corner) does
not show any significant trace of the proton band. The maximum contamination of
the TAPS neutron sample by recoil protons was estimated from these spectra to be
below 3 \% (Monte Carlo simulations indicate a maximum contamination at the 
4\% level). Finally, for the $\eta\pi^o$ final state (upper right corner),
which is not further discussed in the present paper, also a clear band for recoil
deuterons is visible. This band is missing for the $\eta$ final state since coherent
production of single $\eta$ mesons off the deuteron is strongly suppressed
\cite{Weiss_01}.
  
\subsection{Determination of cross sections and systematic uncertainties}
\label{schap:norma}

The extraction of cross sections from the analyzed yields requires several pieces of
information: the $\eta\rightarrow 6\gamma$ decay branching ratio, the target surface 
density, the incident energy dependent photon flux, and the detection efficiency of 
the combined Crystal Barrel - TAPS calorimeter with particle identification detectors 
for photons, protons, and neutrons including all analysis cuts. 

The decay branching ratios for $\eta\rightarrow 3\pi^0$ and $\pi^0\rightarrow 2\gamma$  
are taken from the Particle Data Group \cite{PDG} as (32.56$ \pm$ 0.23)\%
and (98.823$ \pm$ 0.034)\%, respectively, resulting in a total branching ratio of 
31.42 \% with a negligible systematic uncertainty. The target surface density 
(5.3 cm long liquid deuterium target, density $\rho\approx$ 0.169 g/cm$^3$)
was 0.26 deuterons/barn with a systematic uncertainty of $\approx$2\%.

The determination of photon flux and detection efficiency are discussed in detail 
in the following sub-sections.

\subsubsection{Flux normalization}

The data have been taken in five blocks of beam time with different electron beam
energies and settings of linear beam polarization, which we label A, B, C, D, E.
They are summarized in Tab. \ref{tab:beam}. Due to these settings 
the statistical quality of the data is less good above 2 GeV incident photon
energy.

\begin{table}[h]
\begin{center}
\caption{Summary of beam times. $E_{e^-}$: electron beam energy, $E_{\gamma_t}$:
maximum energy of tagged photons, $E_{pol}$: energy of maximum linear photon beam
polarization, $\Phi_o$: energy integrated electron flux on the tagging detectors.
Total life time: beam time multiplied by acquisition life time fraction.} 
\label{tab:beam}       % Give a unique label
% For LaTeX tables use
\begin{tabular}{|c|c|c|c|c|c|}
\hline\noalign{\smallskip}
characteristics & A & B & C & D & E\\
\hline
$E_{e^-}$ [GeV] & 2.6 & 2.6 & 3.2 & 3.2 & 3.2\\
\hline
$E_{\gamma_t}$ [GeV] & 2.0 & 2.0 & 2.5 & 2.5 & 2.5\\
\hline
$E_{pol}$ [GeV]  & 1.0  & 1.0 & unpol. & 1.2 & 1.6 \\
\hline
total life time [h]  & 138 & 18 & 189 & 25 & 25 \\
\hline
$\Phi_o$ [10$^{7}e^-$/s] & 1.75 & 1.6 & 1.6 & 2.8 & 2.8\\ 
\noalign{\smallskip}\hline
\end{tabular}
\end{center}
\end{table}

The photon flux was determined in the following way. The number of scattered electrons
detected by the tagger focal plane counters (scintillating fibers) was determined with
scalers. The scalers were not life time gated, but the experiment dead time (typically
40 \%) was recorded with separate scalers and corrected. Due to an 
electronics/data acquisition problem the scaler information was incorrectly handled 
for part of the data. This resulted in a loss of the absolute normalization for the
beam time blocks B, D, and E, which, however, represent only a minor part of the total
statistics. In order to recover the total available statistics, these three data blocks
were relatively normalized to the data sets A, C, using the total cross sections
for inclusive $\eta$-production and double $\pi^0$ production.

The absolute normalization furthermore requires the measurement of the fraction of
correlated photons that impinge on the target. This was done
with special tagging efficiency runs, where the intensity of the photon beam was
directly measured at the end of the beam line at reduced electron beam intensity.
The average tagging efficiency, which is rather flat as function of beam energy,
was $\approx$95 \% for beam time A and $\approx$75 \% for beam time C (for the 
latter the beam quality was less good).

\begin{figure}[ht]
\resizebox{0.5\textwidth}{!}{%
% normal version
 \includegraphics{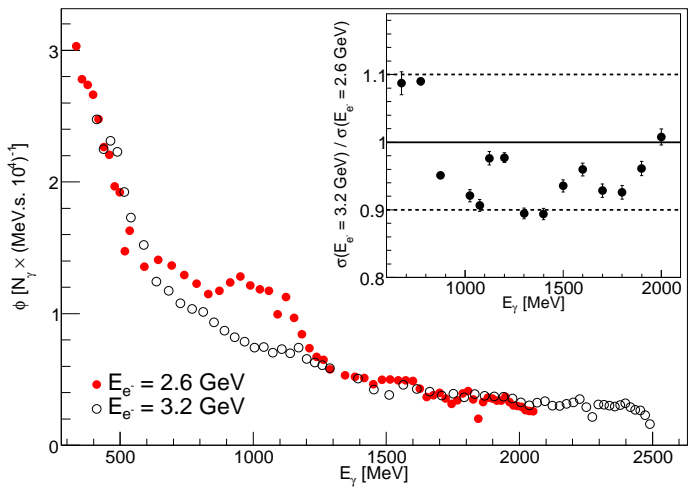}  
}
\caption{Main figure: incident photon flux for beam time A (red dots) and beam time
C (open black circles). Insert: ratio of inclusive $\eta$-photoproduction cross
sections for the two beam times. 
}
\label{fig:flux}       % Give a unique label
\end{figure}

The photon flux, i.e. the product of the spectrum of electron scalers throughout the
tagger detector and the tagging efficiency,
for beam times A and C are compared in Fig. \ref{fig:flux}.
The overall shapes follow
the typical $1/E_{\gamma}$ behavior of bremsstrahlung. For beam time A the peak from
coherent bremsstrahlung is visible around 1 GeV photon energy. The visible structures
are due to systematic effects of individual tagger counters. They are also present in
the extracted yields and cancel in the cross sections. The insert shows
the ratio of the total cross sections for inclusive $\eta$-photoproduction obtained
with the respective photon fluxes for beam times A and C. For most of the energy range
systematic deviations are between 0~\% and -10~\%. At the lowest incident photon
energies they are around +10\%. For the measurement with the 3.2 GeV electron beam, 
this energy region corresponds to the very edge of the tagged range, where systematic
uncertainties tend to be larger. For the final result all data sets were averaged 
according to their statistical weights and a systematic photon flux uncertainty of 
10~\% was estimated.

\subsubsection{Monte Carlo simulations and detection efficiency}

The detection efficiency of the Crystal Barrel/TAPS setup was modelled with Monte 
Carlo simulations based on the GEANT3 package \cite{Brun_86}. The simulations include 
all relevant properties of the calorimeter, including geometrical acceptance, charged
particle identification, trigger efficiency, response of all detector modules, 
and analysis cuts. They include also information about inefficient or malfunctioning 
individual detector modules. This means that the extracted detection efficiencies
are effective ones, which cannot be directly applied to data sets taken under 
different conditions.
As far as photon detection and identification of $\eta$-mesons
via the $\eta\rightarrow 3\pi^0\rightarrow 6\gamma$ decay chain are concerned, some
details are already given in \cite{Mertens_08}, which used an identical setup for
the study of $\eta$-photoproduction from heavy nuclei. However, here in addition the
detection efficiency for recoil nucleons plays a crucial role. The simulations were
done with the GEANT-CALOR program package \cite{Zeitnitz_01}, which is optimized for 
hadronic interactions from the few MeV to several GeV range, including 
the interactions of low energy neutrons. 

\begin{figure}[ht]
\resizebox{0.5\textwidth}{!}{%
% normal version
  \includegraphics{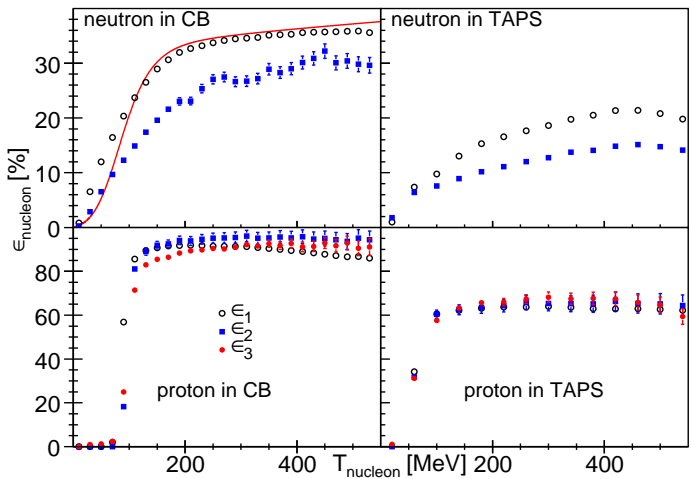}  
}
\caption{Detection efficiency for recoil nucleons. Upper row: neutrons, lower row
protons, left hand side: Crystal Barrel, right hand side: TAPS. For all figures:
$\epsilon_1$ (open black circles): MC simulation for isotropically emitted nucleons
without further detector hits, $\epsilon_2$ (blue squares): simulation with six
decay photons from $\eta$-meson decays, $\epsilon_3$ (red dots): experimentally 
determined proton detection efficiency from $\eta$ and $2\pi^0$ photoproduction from 
proton (hydrogen) target (see text).
Solid (red)line: measured neutron detection efficiency for CB at CERN 
\cite{Schaefer_93}.
}
\label{fig:neffi}       % Give a unique label
\end{figure}

The results for the recoil nucleon detection efficiencies are summarized in Fig.
\ref{fig:neffi}. For both, neutrons and protons, in the first step isotropically 
distributed nucleons were simulated. The detection efficiency was determined as 
function of kinetic energy $T$ and laboratory polar angle $\Theta$.
As expected, angular dependencies of the efficiency (not shown) are negligible for 
the CB apart from the very edge of the detector. They are more important in the 
TAPS detector (changing angle of incidence towards the outer edge, increasing 
discriminator thresholds towards the beam pipe). Fig. \ref{fig:neffi} 
(black open circles) shows the kinetic energy dependence of the detection efficiency
obtained with this simulation, averaged over the polar angle range of the detectors.

In case of the proton, the simulation can be checked with data from the liquid 
hydrogen target measured with the same setup. For this purpose, the reactions
$\gamma p\rightarrow p\eta\rightarrow p6\gamma$ and 
$\gamma p\rightarrow p\pi^0\pi^0\rightarrow p4\gamma$ were analyzed using similar
procedures for invariant and missing mass analyses as discussed above. The proton
detection efficiency $\epsilon_p(T,\Theta)$ is then simply given as the ratio of
events with detected proton to all events of the respective reaction. The result,
again averaged over the polar angles, agrees quite well with the simulation
(see Fig. \ref{fig:neffi}, red dots). In case of the neutron, experimental
information comes from the measurement of the neutron detection efficiency of the CB
when it was installed at LEAR at CERN \cite{Schaefer_93}. Therefore, in this 
case the simulation was done with the parameters (e.g. thresholds) characterizing
the CB setup at CERN. The results are also compared in Fig. \ref{fig:neffi}. 
Agreement is quite good at larger $T$. Some discrepancies are visible at low $T$,
however, here the result is strongly dependent on detector thresholds and neutron
energy calibration, which may not have been reproduced perfectly in the present
simulation of the old CB setup.
In case of TAPS the neutron detection efficiency had been measured in an experiment 
at the Mainz MAMI accelerator using the $\gamma p\rightarrow n\pi^0\pi^+$ reaction 
for kinetic energies below 250 MeV \cite{Hejny_99}. The results are consistent with 
the simulation when the conditions of the Mainz setup are used (at 250 MeV simulated:
$\epsilon_1\approx$ 18.5\%, from data: $\epsilon_1\approx$ 19.1\%).  

Finally, it must be considered that neutrons from the $\gamma n\rightarrow n\eta$ 
reaction are identified in this experiment in events with seven neutral hits
by first assigning six hits via the invariant mass analysis to the
$\eta\rightarrow 3\pi^0\rightarrow 6\gamma$ decay chain. This introduces efficiency
losses due to combinatorial background. Therefore, a simulation was done, where
events from $\gamma n\rightarrow n\eta$ were produced with a phase space event 
generator. The analysis then mimicked the whole reconstruction process, including 
the identification of the neutron out of the seven neutral hits. The resulting
`effective' neutron detection efficiency is also shown in Fig. \ref{fig:neffi}
(blue squares). It is significantly lower than the `raw' neutron efficiency.
The same kind of analysis was also done for the proton case. However, due to the
additional information from Inner- and TAPS-veto detectors, in this case the 
effects are small. As final result of this analysis, recoil detection efficiencies 
for protons and neutrons are available as function of laboratory polar angle and 
kinetic energy. 

\begin{figure}[ht]
\resizebox{0.5\textwidth}{!}{%
% normal version
 \includegraphics{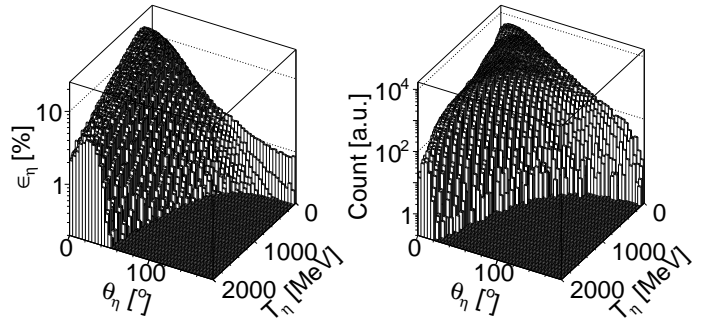}  
}
\caption{Left hand side: simulated detection efficiency
(including trigger efficiency) for $\eta\rightarrow 6\gamma$
as function of $\eta$ laboratory kinetic energy and polar angle. Right hand
side: distribution of observed $\eta$ mesons in dependence on the same parameters. 
}
\label{fig:geffi}       % Give a unique label
\end{figure}

The detection efficiency for $\eta$-mesons has been simulated as function of their
laboratory polar angles and kinetic energies in the same way as discussed in detail 
in \cite{Mertens_08}. The result is shown in Fig. \ref{fig:geffi} (left hand side)
and compared to the distribution of the experimentally detected $\eta$-mesons (Fig.
\ref{fig:geffi}, right hand side). The branching ratio of the 
$\eta\rightarrow 6\gamma$ decay is not included in this efficiency (i.e. shown is the
efficiency to detect $\eta$-mesons which decay with 100~\% branching ratio into six
photons). The acceptance 
of the detector covers the full phase space of the reaction. However, the absolute 
values of the efficiency are small in particular for mesons at backward angles. 
This is due to the efficiency of the trigger, which was only based on photon 
detection in TAPS. Note that the recoil nucleon detection efficiencies discussed 
above do not include trigger efficiencies, since the meson is assumed to trigger. 
 
The total detection efficiencies for the inclusive channel (no recoil nucleons 
required), the reactions in coincidence with recoil protons, and in coincidence 
with recoil neutrons were then calculated in two different ways to estimate 
systematic effects.

\begin{figure}[tht]
\resizebox{0.48\textwidth}{!}{%
% normal version
 \includegraphics{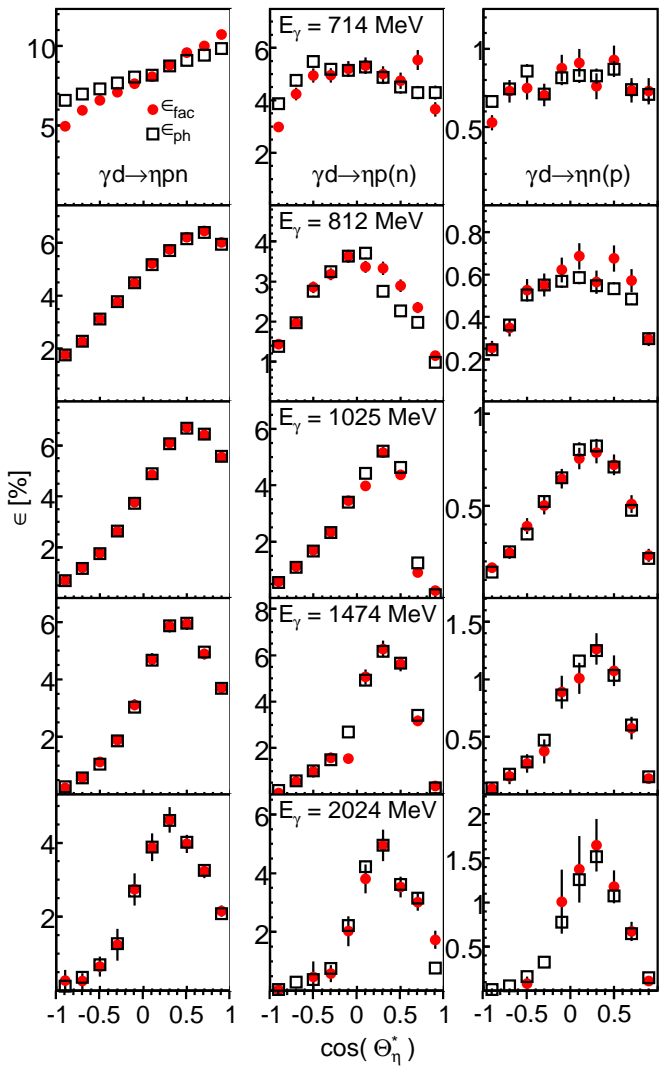}  
}
\caption{Total detection efficiencies. 
Left column: $\gamma d\rightarrow \eta np$ (inclusive reaction), 
center column: $\gamma d\rightarrow \eta p(n)$ (recoil protons detected), 
right column $\gamma d\rightarrow \eta n(p)$ (recoil neutrons detected) 
for different incident photon energies as function of the $\eta$ cm polar angle. 
(Red) dots: efficiency from quasi-factorization ($\epsilon_{fac}$), 
(black) open squares: efficiency from phase-space simulation( $\epsilon_{ph}$).
}
\label{fig:deffi}       % Give a unique label
\end{figure}

In the first approach, $\eta$-photoproduction off quasi-free nucleons was simulated 
with a phase-space event generator, taking into account the effects of nuclear Fermi 
smearing. The simulated events were analyzed in the same way as the data and the 
detection efficiency $\epsilon_{ph}$ was calculated from the ratio of simulated and 
detected events as function of incident photon energy and meson cm polar angle. 
The results are shown for some ranges of incident photon energy in 
Fig.~\ref{fig:deffi} as open squares. This simulation is in so far model dependent 
as the correlation between meson energies and angles as well as the correlation 
between mesons and recoil nucleons relies on the phase-space assumption. Possible 
deviations might occur in the three-body final state of meson, participant, and 
spectator nucleon for example due to final state interactions.

In the second approach, the above discussed distributions of the detection 
efficiencies for mesons and recoil nucleons as function of particle laboratory angle 
and kinetic energy were applied event-by-event to the data. This step is completely 
model independent since the correction is only based on measured quantities
(recoil nucleon energies from time-of-flight in TAPS and from reaction kinematics in
CB). 
However, it does not correct for the missing mass cuts (roughly a factor of two for 
the cut at $\Delta m <$ 0 MeV). The loss factor due to this cut was again simulated with 
an event generator based on reaction phase space and corrected. The results for this 
almost model independent efficiencies $\epsilon_{fac}$ from the quasi-factorization 
are shown as red dots in Fig. \ref{fig:deffi}. 
Obviously, the results obtained with both methods are in quite good agreement.

As expected, due to the trigger efficiency, for all reaction channels the detection 
efficiency is small at backward angles, in particular at higher incident photon 
energies. For the exclusive reactions, in particular the proton channel, it is also 
small for meson forward angles, corresponding to recoil nucleons emitted at large 
angles with small kinetic laboratory energies. 

\begin{figure}[ht]
\resizebox{0.5\textwidth}{!}{%
% normal version
 \includegraphics{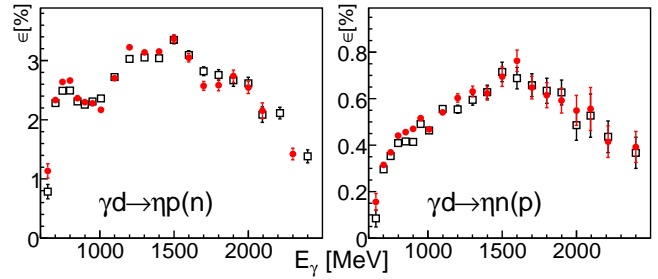}  
}
\caption{Angle integrated total detection efficiencies $\epsilon_{ph}$ (open squares) 
and $\epsilon_{fac}$ as function of incident photon energy. 
}
\label{fig:teffi}       % Give a unique label
\end{figure}

The angle integrated detection efficiencies as function of incident photon energy 
are shown in Fig. \ref{fig:teffi}. The curve for the proton channel shows some 
structure at photon energies of 1 GeV. This can be explained with 
Figs. \ref{fig:regions},~\ref{fig:neffi}. At the lowest incident photon energies, all
recoil protons are detected in TAPS. Around 900 MeV, part of the protons reaches the
Barrel. Since the detection threshold in the Barrel is higher than in TAPS, this 
leads to a decrease of the efficiency. But since proton detection above the threshold
was more efficient in the Barrel ($\approx$90\%) than in TAPS ($\approx$60\%) the 
overall efficiency rises again for higher incident photon energies. Since the 
neutron detection efficiency varies smoothly for both detectors no structure is 
present for quasi-free production off the neutron.

\subsubsection{Summary of systematic uncertainties}

Typical statistical uncertainties for the total cross sections range from 
0.5~\% - 5~\% for the inclusive data, from 1~\% - 10~\% for the data with coincident
protons and from 2~\% - 20\% for the data with coincident neutrons (the first number
corresponds to the range of the S$_{11}$(1535) peak maximum, the second number to
maximum incident photon energies). As discussed below, systematic uncertainties 
are of comparable size for high incident photon energies but dominate in the 
S$_{11}$ range.

Systematic uncertainties are in three different categories: overall uncertainties 
which cancel exactly in the comparison of different reaction channels, uncertainties 
which are similar for different reaction channels and cancel to a large extent in 
ratios, and reaction channel related uncertainties which do not cancel.

In the first category all cross sections are subject to an overall systematic 
uncertainty of the photon flux of $\approx$10~\%. The overall uncertainty of the 
target thickness of a few per cent is comparably small and the uncertainty of the 
decay branching ratio (below 1 \%) is negligible.

The second class of uncertainties is related to the detection of the $\eta$-mesons.
The main steps of the analysis, which have to be reproduced by the Monte Carlo 
simulations, are the detection, identification, and calibration of photon showers, 
the invariant mass analysis for the identification of the mesons, and the missing 
mass analysis removing background from reactions with additional mesons in the final 
state. 

At the most basic level, a stringent limit for systematic uncertainties 
arising from the detection of photons and the subsequent identification of mesons 
via the invariant mass analysis can be derived from a comparison of the results for 
$\eta$-photoproduction off the free proton. This has been previously analyzed for 
the same setup for the $\eta\rightarrow 2\gamma$ and the $\eta\rightarrow 6\gamma$ 
decay channels \cite{Crede_05,Bartholomy_07,Crede_09}. Systematic effects would enter
cubed into the six-photon channel, but agreement was found on average at the 2~\% 
level. Furthermore, also the present simulations with the two different methods 
- one relying on the phase-space event generator, the other using the 
quasi-factorization of the detection efficiency, are in good agreement. From this 
we estimate a typical 5~\% uncertainty for a successful $\eta$-reconstruction. 
The uncertainty related to the separation of signal and background in the invariant 
mass spectra (fitted with line shape and background polynomial) is not explicitly 
treated as an additional systematic effect, but the fit uncertainty is included into 
the statistical errors. 

Finally, the missing mass cut deserves special attention 
(see Fig. \ref{fig:mima}). At incident photon energies below 0.8 GeV, the spectra are 
practically background free. At higher energies background from $\eta\pi$ final 
states and the tail of the missing mass distribution for single $\eta$-production 
arising from the momentum distribution of the bound nucleons mix. The simulations 
indicate that the background reactions do not contribute in the region of negative 
missing masses (cf. Fig. \ref{fig:mima}). Since only those events 
were accepted, background contamination is estimated at most at the per cent level. 
However, due to this cut, the simulation must closely reproduce the shape of the 
missing mass peak, including effects of Fermi motion. This is the case for the 
background free peaks at low incident photon energies and for all energies for the 
peak shape at the non-contaminated side. In the background contaminated regions 
the data can be reproduced by a summation of the simulated structures for peak and 
background (with properly adjusted relative contributions), but no direct check is 
possible. From the agreement between data and simulations and the fraction of the 
missing mass signal extending into the tail region we estimate for the missing mass
analysis a systematic uncertainty of 3~\% for the total cross section in the 
threshold region rising to 15~\% at 2.5 GeV. Altogether, independent on the 
reaction channel we estimate the systematic uncertainty for the $\eta$ detection 
at 8~\% at production threshold rising to 20~\% at 2.5 GeV.           

\begin{table}[h]
\begin{center}
\caption{Summary of systematic uncertainties for the quasi-free reactions.
$^{1)}$ photon flux, target thickness,
decay branching ratios, $^{2)}$ trigger efficiency, $\eta$ analysis
cuts, $\eta$ detection efficiency. When two numbers are given the first corresponds
to the threshold energy range, the second to $E_{\gamma}$= 2.5 GeV, 
and linear interpolation is used in between.} 
\label{tab:systematics}       % Give a unique label
% For LaTeX tables use
\begin{tabular}{|c|c|c|}
\hline\noalign{\smallskip}
source & $\gamma d\rightarrow (n)p\eta $  & $\gamma d\rightarrow (p)n\eta$ \\
\hline
overall normalization$^{1)}$ & 10~\% & 10 \% \\
\hline
$\eta $ detection$^{2)}$ & 8~\% - 20~\%  & 8~\% - 20~\%\\
\hline
recoil nucleon detection  & 8~\%  & 15~\%  \\
\noalign{\smallskip}\hline
\end{tabular}
\end{center}
\end{table}

The most critical uncertainties are related to the detection of the recoil nucleons.
From the agreement between simulated and measured efficiencies we estimate on average 
$\approx$8~\% uncertainty for recoil protons and 15~\% for recoil neutrons.
Uncertainties can be larger (in particular for the proton) for kinematical parameters
where they are partly detected in TAPS and partly in the CB, i.e. for incident
photon energies from 0.8 GeV - 0.9 GeV and cos($\Theta_{\eta}^{\star}$)$>$ 0 and
at higher incident photon energies for cos($\Theta_{\eta}^{\star}$)$\approx$ -0.1
(cf. Fig. \ref{fig:regions}). The above discussed systematic uncertainties are
summarized in Tab. \ref{tab:systematics}.

\begin{figure}[ht]
\resizebox{0.5\textwidth}{!}{%
% normal version
  \includegraphics{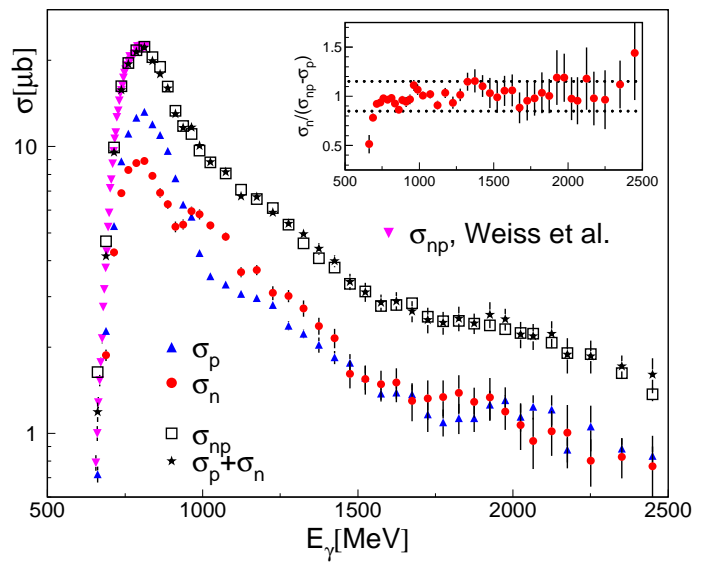}  
}
\caption{Comparison of total cross sections. 
(Blue) upward triangles: quasi-free proton cross section $\sigma_p$, 
(red) dots: quasi-free neutron cross section $\sigma_n$, (black)
open squares: inclusive quasi-free cross section $\sigma_{np}$, 
(black) stars: $\sigma_n+\sigma_p$. 
Downward (magenta) triangles: inclusive quasi-free cross 
section from Weiss et al. \cite{Weiss_03}. Insert: ratio of neutron cross sections.}
\label{fig:pin}       % Give a unique label
\end{figure}

\begin{figure}[ht]
\resizebox{0.47\textwidth}{!}{%
% normal version
  \includegraphics{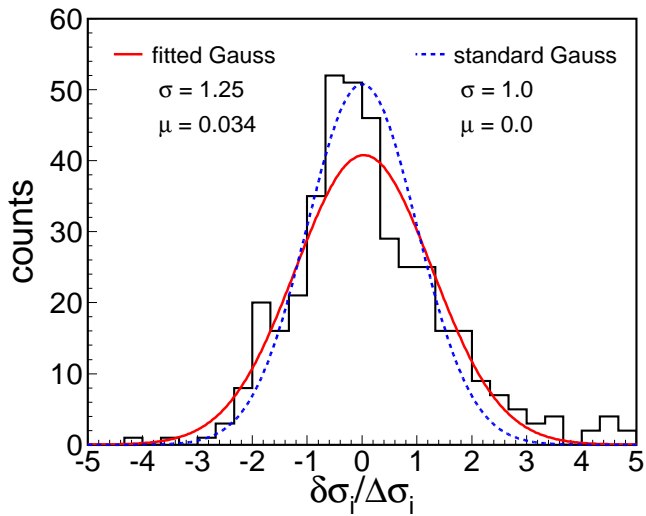}  
}
\caption{Distribution of deviations between $d\sigma_n/d\Omega$ and
$d\sigma'_n/d\Omega = d\sigma_{np}/d\Omega-d\sigma_{p}/d\Omega$.
Solid (red) curve: fitted Gaussian distribution (width $\sigma$=1.25, 
mean $\mu$=0.034), dashed (blue) curve: standard Gauss: ($\sigma$=1, $\mu$=0)}
\label{fig:statis}       % Give a unique label
\end{figure}

As already discussed in \cite{Jaegle_08,Jaegle_11} the nucleon detection uncertainty 
may be checked in an independent way. The cross section of the coherent process 
$\gamma d\rightarrow d\eta$ is negligible \cite{Weiss_01} compared to the quasi-free 
reaction. Therefore, the quasi-free reactions $\sigma_{np}$ (inclusive, no condition
for recoil nucleons), $\sigma_{p}$ (coincident recoil protons), and $\sigma_n$
(coincident recoil neutrons) must obey 
$\sigma_{np} = \sigma_p + \sigma_n$. This is demonstrated in Fig. \ref{fig:pin}, 
where the sum of the quasi-free proton and neutron cross sections is compared to 
the inclusive cross section. The agreement is excellent and allows an independent
extraction of the neutron cross section as $\sigma'_n = \sigma_{np}-\sigma_{p}$. 
The ratio $\sigma_n/\sigma'_n$ of the two results is shown in the insert of 
Fig.~\ref{fig:pin}. The agreement is within statistical uncertainties for most data
points and typical deviations do not exceed the 15~\% level. As a further test,
the distribution of the deviations $\delta\sigma_{i}$ normalized by the statistical
uncertainties $\Delta\sigma_{i}$    
\begin{equation}
\frac{\delta\sigma_i}{\Delta\sigma_i}\equiv 
\frac{d\sigma'_n/d\Omega-d\sigma_n/d\Omega}
{\sqrt{\Delta^2 (d\sigma'_n/d\Omega)+\Delta^2 (d\sigma_n/d\Omega)}}
\end{equation} 
for all data points (420 entries) of the angular distributions from production 
threshold to 2.5 GeV is compared to a Gaussian distribution in Fig. \ref{fig:statis}.  
The fitted Gaussian distribution corresponds to a width of $\sigma=(1.25\pm0.10)$ 
and a mean of $\mu=(0.034\pm 0.110)$, fairly close to a standard Gaussian distribution.
In particular, the mean is not significantly different from zero so that no 
indication for a systematic deviation is indicated. 

This is a very stringent test for systematic uncertainties related to the recoil 
nucleon detection since it is the neutron detection efficiency which enters in the 
extraction of $\sigma_n$, while only the inherently different proton detection 
efficiency enters into $\sigma'_n$. The result means that the corresponding 
systematic uncertainties quoted in Tab \ref{tab:systematics} derived from the 
analysis of the recoil nucleon detection efficiencies are probably pessimistic.

Due to this good agreement total neutron cross sections as function of incident photon 
energy are given in this paper as weighted averages $\langle\sigma_n\rangle$ of 
$\sigma_n$ and  $\sigma'_n$ which improves the statistical quality. For the shape of
the angular distributions only the direct measurement with the neutron coincidence 
is used since due to the small detection efficiency at extreme angles (in particular
for the proton) systematic effects for $\sigma'_n$ are larger.   

\section{Results and discussion}
\label{results}

Throughout this paper all quasi-free differential cross sections are given in the cm 
(center-of-momentum) system of the incident photon and a target nucleon {\it at rest}. 
This simplifies the comparison to angular distributions measured off the free nucleon
since apart from the immediate threshold region such quasi-free cross sections are only
moderately smeared out by Fermi motion in this system, while in the $\gamma d$ system
they have completely different shapes due to the Lorentz boosts
(see \cite{Krusche_95a} for details).

\subsection{Comparison to previous results}

\begin{figure}[bht]
\resizebox{0.5\textwidth}{!}{%
% normal version
  \includegraphics{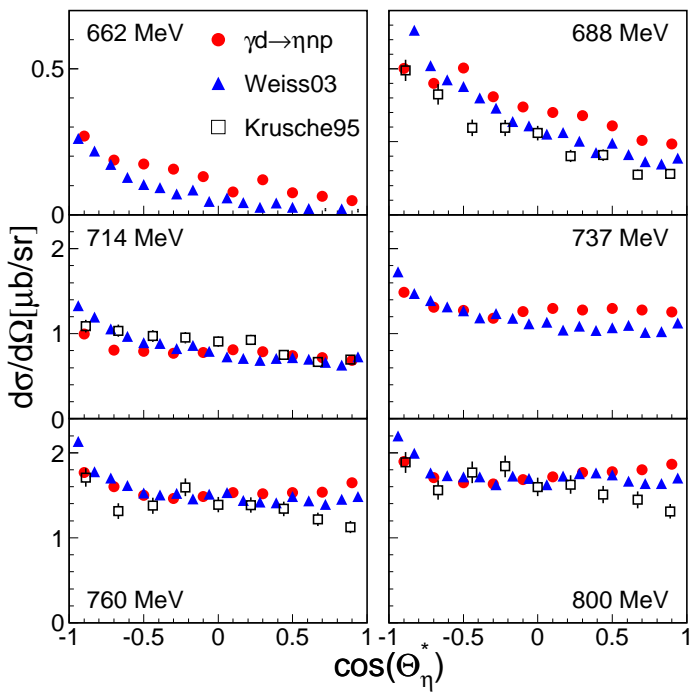}  
}
\caption{Comparison of the inclusive cross section $\sigma_{np}$ to previous results 
(Krusche95 \cite{Krusche_95a}, Weiss03 \cite{Weiss_03}). Errors are only statistical.
}
\label{fig:weiss}       % Give a unique label
\end{figure}

The inclusive cross section $\sigma_{np}$ has been previously measured twice 
\cite{Krusche_95a,Weiss_03} at the Mainz MAMI accelerator with different
configurations of the TAPS detector (the setup used for \cite{Weiss_03} covered
a larger solid angle than for \cite{Krusche_95a}) up to incident photon energies of 
0.8 GeV. The total cross section from \cite{Weiss_03} is included in 
Fig.~\ref{fig:pin}. Typical angular distributions are compared in 
Fig.~\ref{fig:weiss}. 

The shapes of the angular distributions are similar for all three experiments.
There is a systematic deviation between the two previous measurements with the 
present data for the absolute scale of the two lowest energy bins. 
These are, however, probably not due to the normalization of the cross section
data but to the systematic uncertainty in the measurement of the incident photon
energy. Both bins are located in the extremely sharp rise of the cross section close 
to threshold (cf. Fig.~\ref{fig:pin}).
In this range, already small effects in the determination of the incident photon beam 
energy are strongly amplified in the magnitude of the cross section. 
The resolution for the incident photon energy in the threshold region was better for 
the previous MAMI experiments, which aimed at very precise threshold measurements in 
view of FSI effects. The MAMI data had typically 2 MeV bin width for the incident 
photon energy with an absolute calibration uncertainty of less than 1 MeV while the 
present data were measured with 10 MeV bin width and a calibration uncertainty of 
several MeV. Therefore, the systematic quality of the previous data in the 
immediate threshold region is almost certainly superior. 
However, this region is not of much interest for the present experiment. 
Agreement is much better and within systematic uncertainties at higher incident 
photon energies. In this range, the 
data from \cite{Krusche_95a} have somewhat larger systematic uncertainties than 
the other two data sets due to the restricted solid angle coverage.

\begin{figure}[th]
\resizebox{0.5\textwidth}{!}{%
% normal version
  \includegraphics{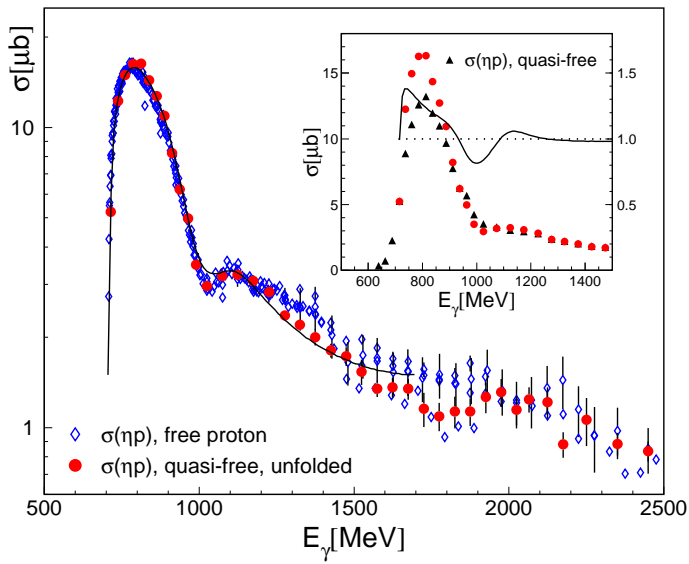}  
}
\caption{Comparison of the total cross section of quasi-free production
off the bound proton to free proton data. Filled (red) circles: quasi-free
$\sigma_p$ corrected for effects of Fermi motion (see text). 
Open (blue) diamonds: free proton data from
\cite{Krusche_95,Renard_02,Dugger_02,Crede_05,Crede_09,Williams_09,McNicoll_10}.
Total cross sections for the free proton data from \cite{Dugger_02,Williams_09} 
have been estimated from the published differential cross sections.
Solid line: eta-Maid model \cite{Chiang_02}.
Insert: comparison of quasi-free proton data (black triangles) to quasi-free
proton data after correction of Fermi motion effects (red filled circles)
(see text). Solid curve: ratio of free and folded cross section (scale at right 
hand side).
}
\label{fig:compa}       % Give a unique label
\end{figure}

The quasi-free proton data are compared in 
Figs. \ref{fig:compa},\ref{fig:p_diff} to free proton data. The total cross
section for the quasi-free reaction has been corrected for the effects of
Fermi motion in the following way. The well known energy dependence of the 
cross section for $\eta$-production off the free proton
was folded with the momentum distribution of the bound proton using the
deuteron wave function in momentum space from \cite{Lacombe_81} as described in
\cite{Krusche_95a}. The ratios of free and folded cross section (solid line in 
the insert of Fig. \ref{fig:compa}) were then applied as correction factors
to the measured quasi-free cross section. Measured and corrected quasi-free 
cross sections are compared in the insert of Fig. \ref{fig:compa}. The
corrected quasi-free cross section is then compared in Fig. \ref{fig:compa}
to the data base of free proton results. The agreement between the free and
quasi-free data is excellent for most incident photon energies. This indicates
that no nuclear effects (FSI, re-scattering) except nuclear Fermi motion
influence the quasi-free data. The latter effects are significant in the steep
slopes of the cross section for photon energies below 1.1~GeV, they are negligible
in the flat region at higher incident photon energies.

Angular distributions for the free and quasi-free $\gamma p\rightarrow p\eta$
reactions are summarized in Fig. \ref{fig:p_diff}. The quasi-free data have not
been corrected for Fermi motion. They have been fitted with Legendre polynomials
\begin{equation}
\frac{d\sigma}{d\Omega} = \frac{q_{\eta}^{\star}}{k_{\gamma}^{\star}}
\sum_{i=0}^{3} A_iP_i(cos(\Theta^{\star}_{\eta})) \ ,
\label{eq:legendre}
\end{equation}
where the $A_i$ are expansion coefficients (higher order coefficients were not
significant). The phase-space factor 
${q_{\eta}^{\star}}/{k_{\gamma}^{\star}}$ (${q_{\eta}^{\star}}$, ${k_{\gamma}^{\star}}$:
meson and photon cm momenta)
is evaluated for the photon - nucleon-at-rest cm system.  

At incident photon energies above 1.1~GeV the angular
distributions are in excellent agreement with the quasi-free data.
Close to threshold the large influence of Fermi motion is visible,
but the comparison of the quasi-free data to the MAID model result
folded with Fermi motion (dashed lines) demonstrates that this 
effect is well under control. A large deviation between free and 
quasi-free data occurs also for incident photon energies around 
975 MeV. In this region the pronounced `dip' in the total free cross
section is filled in by Fermi motion from the tail of the S$_{11}$
resonance, however again the folded cross section is in good agreement
with experiment. 

In summary, we conclude that after correction for the effects of nuclear
Fermi motion the absolute scale and the shape of the angular distributions
of the quasi-free proton data agree very well with the most recent and
most precise measurements of $\eta$-photoproduction from the free proton.
This is the systematic basis for the discussion of the quasi-free neutron
data.

\subsection{The quasi-free reaction {\boldmath{$\gamma n\rightarrow n\eta$}}
off the neutron}

\begin{figure}[th]
\centerline{\resizebox{0.45\textwidth}{!}{%
% normal version
  \includegraphics{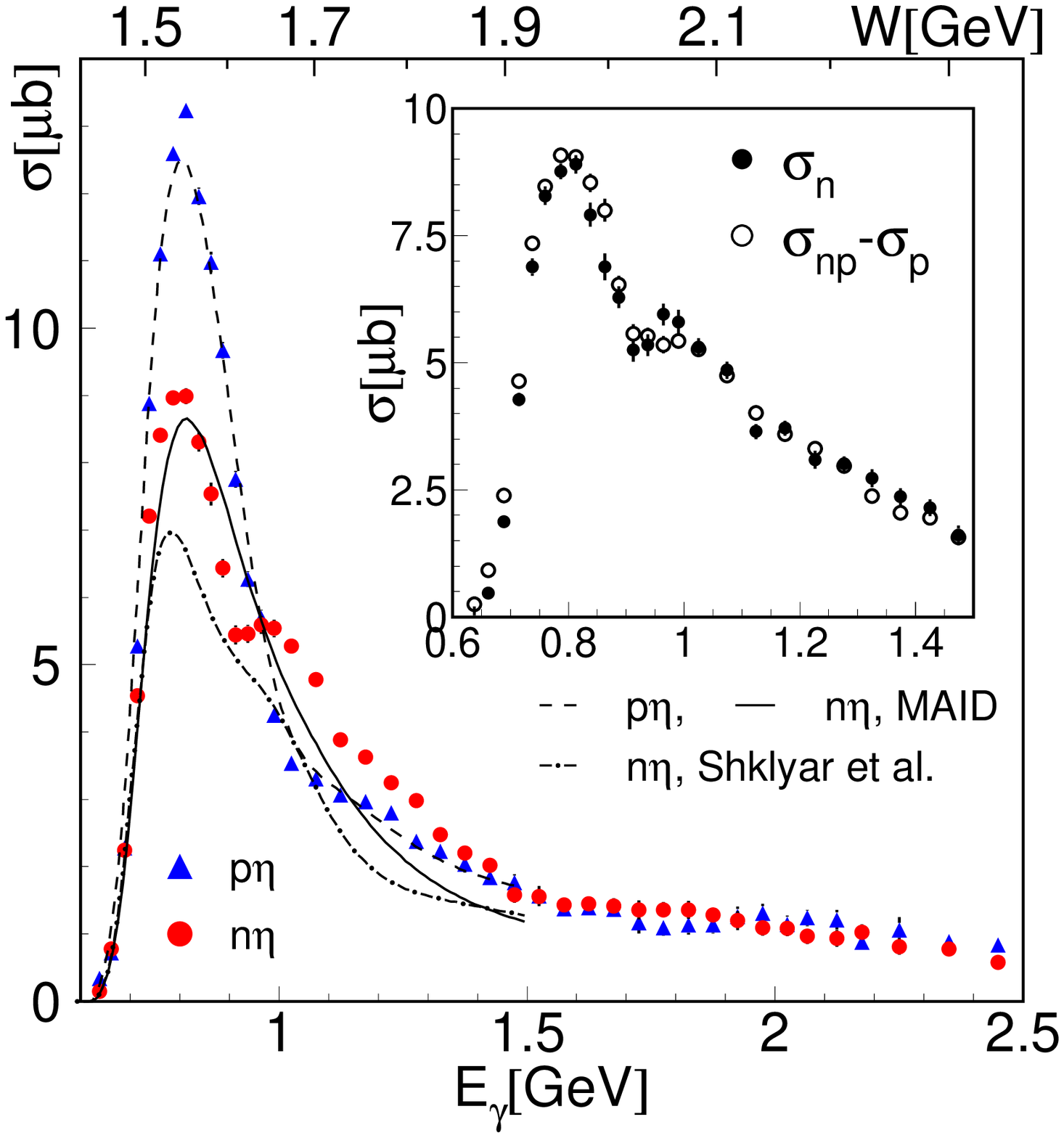}}}

\vspace*{0.5cm}  
\centerline{\resizebox{0.45\textwidth}{!}{%  
\includegraphics{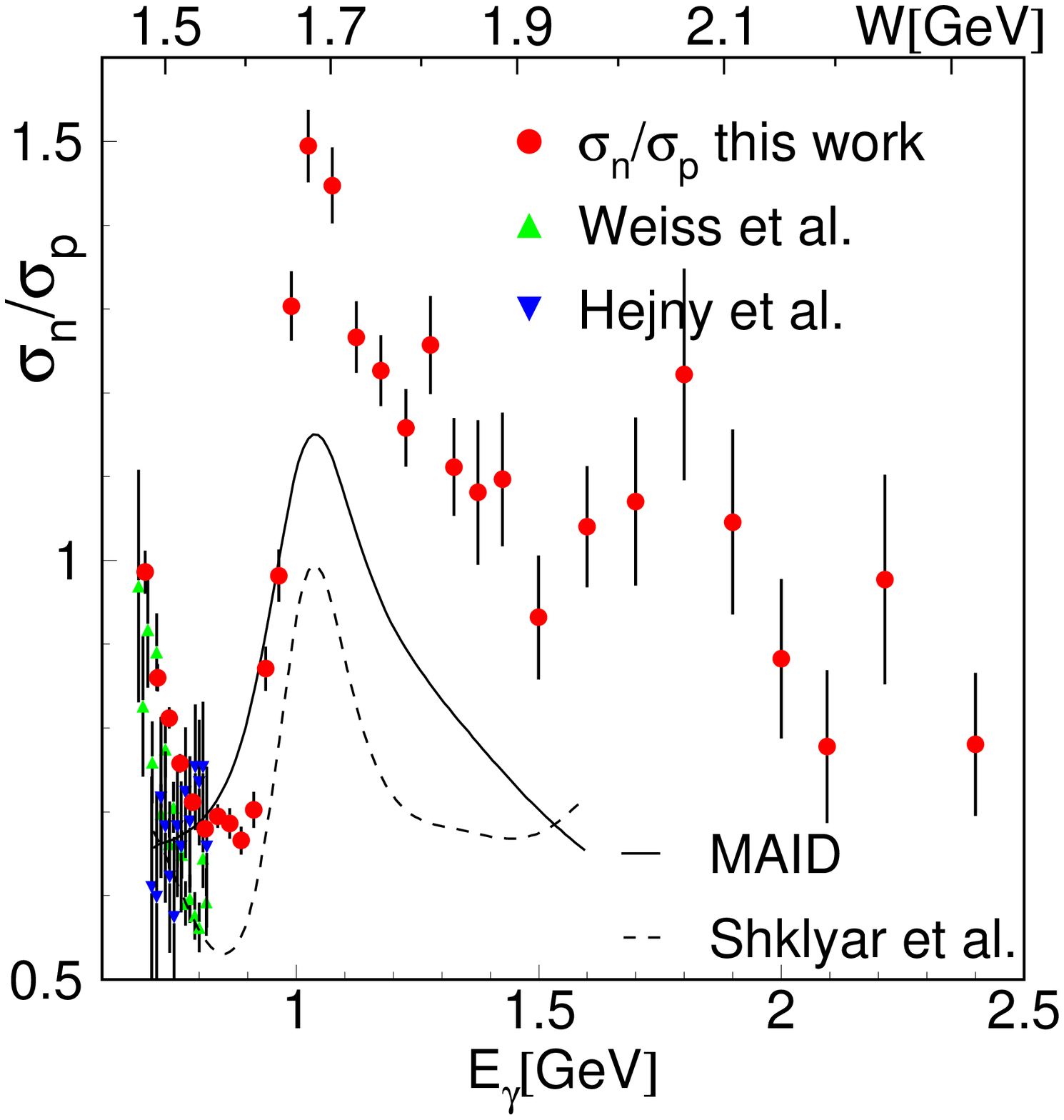}}     
}
\caption{Comparison of quasi-free proton and neutron excitation function.
Upper part:
Curves model results, dashed: Eta-MAID for proton \cite{Chiang_02}, 
solid: Eta-MAID for neutron \cite{Chiang_02}, 
dash-dotted: Shklyar et al. \cite{Shklyar_07}. Insert: comparison of the 
total neutron cross sections extracted from the coincident measurement of
neutrons ($\sigma_n$) and the difference of inclusive and proton data ($\sigma_n'$).
Bottom part:
Cross section ratio $\sigma_n/\sigma_p$ compared to previous data
from quasi-free production off the deuteron \cite{Weiss_03} and off $^4$He
\cite{Hejny_99} and model results (solid: MAID \cite{Chiang_02}, 
dashed: Shklyar et al. \cite{Shklyar_07}) folded with Fermi motion.
}
\label{fig:ntot}       % Give a unique label
\end{figure}

The total cross section of the quasi-free reaction off the neutron is compared 
to the quasi-free proton data in Fig. \ref{fig:ntot}, which also shows the
neutron/proton cross section ratio. The behavior at low incident photon
energies ($E_{\gamma}\leq$800 MeV) is consistent with previous results.
The cross section ratio in the S$_{11}$-region is close to 2/3 and rises to
the kinematic threshold close to unity because in the immediate vicinity
of the threshold the participant - spectator approach becomes meaningless
(dictated by energy and momentum conservation at threshold `participant' and 
`spectator' nucleon have always identical momenta). 

\begin{figure*}[th]
\resizebox{0.99\textwidth}{!}{%
% normal version
  \includegraphics{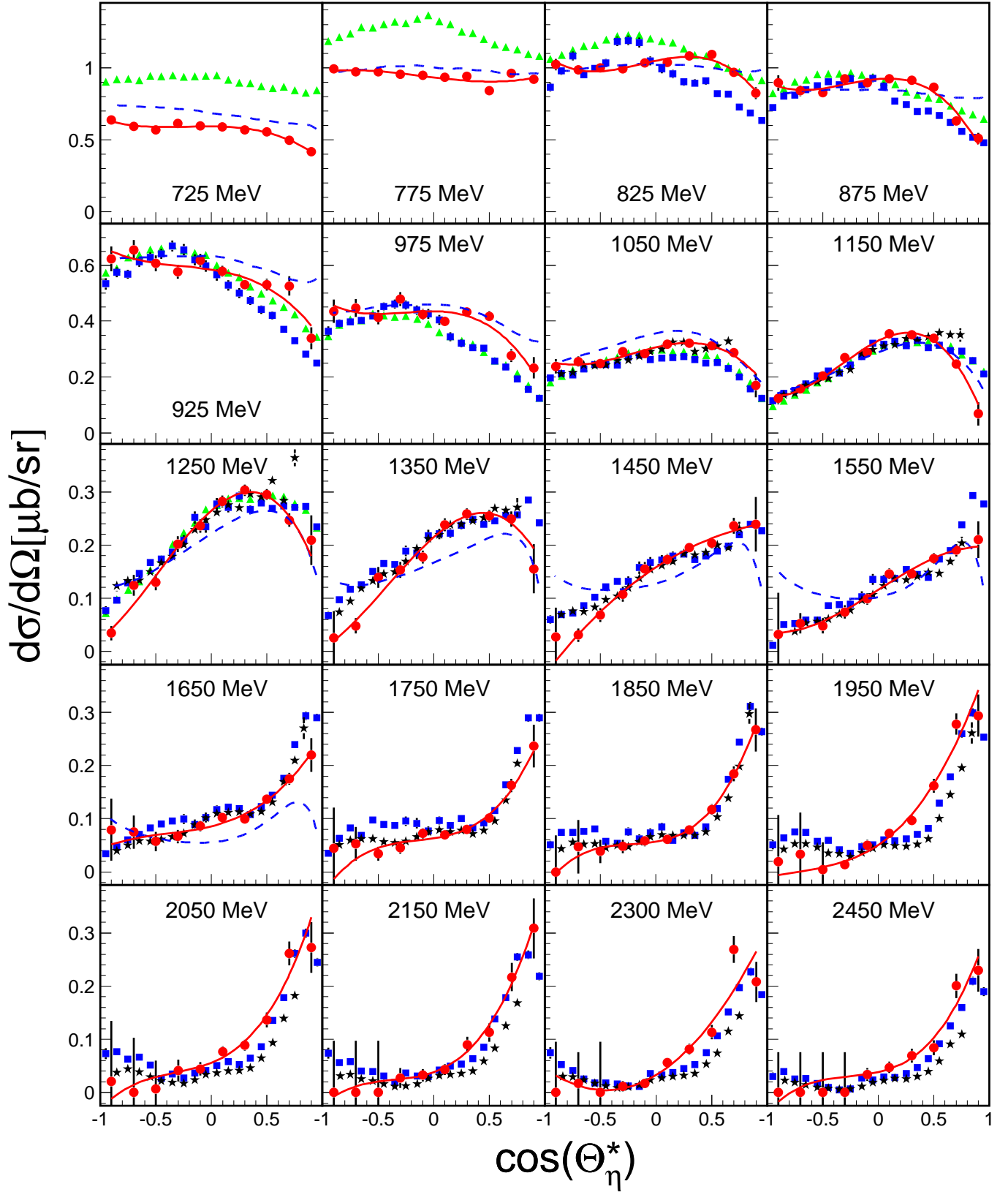}  
}
\caption{Differential cross sections for $\gamma p\rightarrow p\eta$.
(Red) filled circles: quasi-free data from present experiment (not corrected
for Fermi motion), (Red) solid lines: fits with Legendre polynomials.
Other symbols free proton data: (blue) squares: \cite{Crede_09},
(black) stars: \cite{Williams_09}, (green) triangles: \cite{McNicoll_10}
(data have been partly re-binned to cover the same energy ranges).
Dashed (blue) lines: Eta-Maid model \cite{Chiang_02} folded with Fermi 
motion}
\label{fig:p_diff}       % Give a unique label
\end{figure*}
\clearpage

\begin{figure*}[th]
\resizebox{0.99\textwidth}{!}{%
% normal version
  \includegraphics{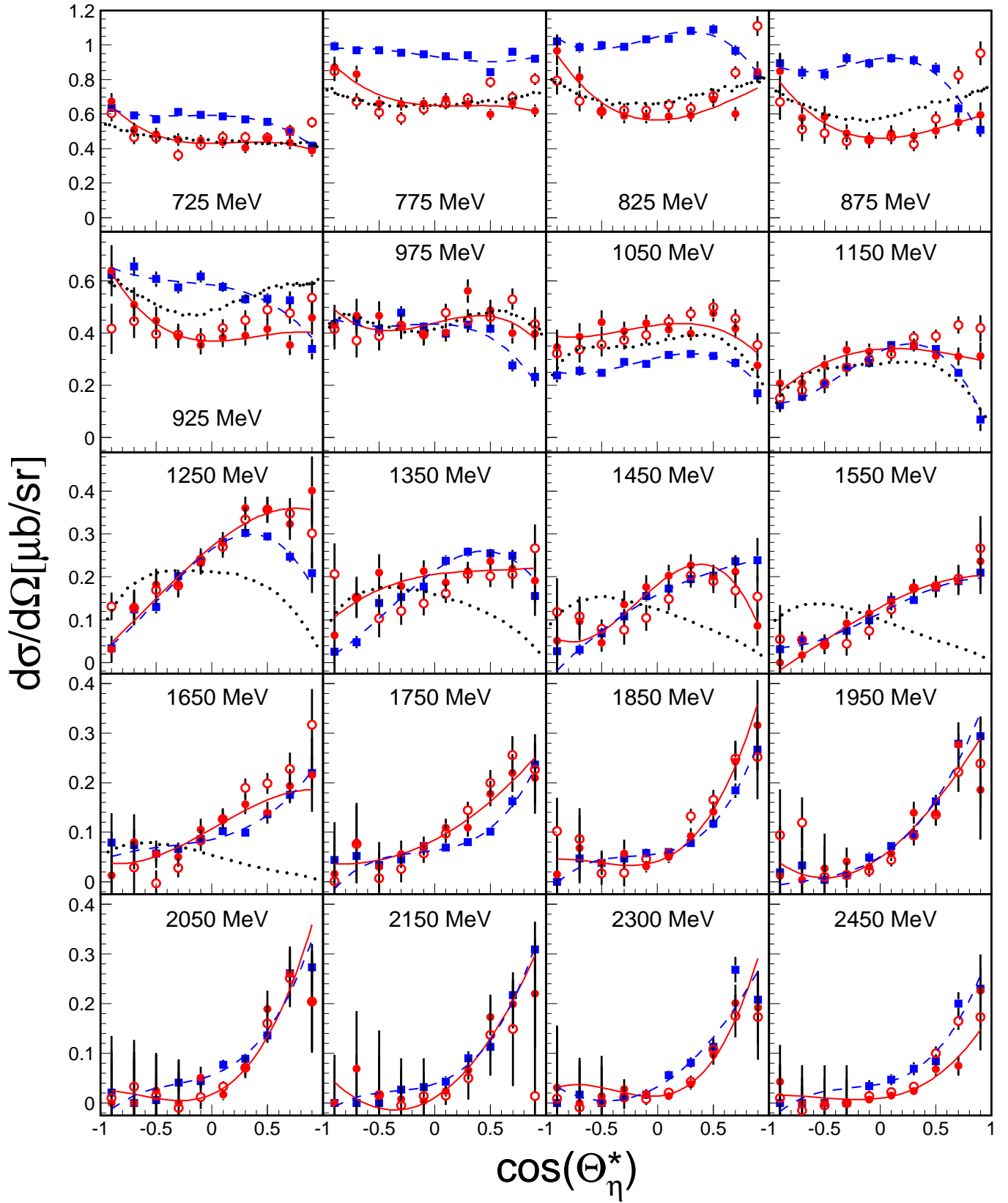}  
}
\caption{Quasi-free angular distributions, labels indicate incident photon energy. 
(Blue) squares: proton coincidence $\sigma_p$, 
(red) filled circles: neutron coincidence $\sigma_n$, 
(red) open circles: difference of inclusive and proton $\sigma'_n$.
Dashed (blue) curves: fit of proton data, 
solid (red) curves: fit of neutron data,
dotted (black) curves: Eta-MAID for neutron folded with Fermi motion.
}
\label{fig:ndiff}       % Give a unique label
\end{figure*}

\begin{figure}[th]
\resizebox{0.5\textwidth}{!}{%
% normal version
  \includegraphics{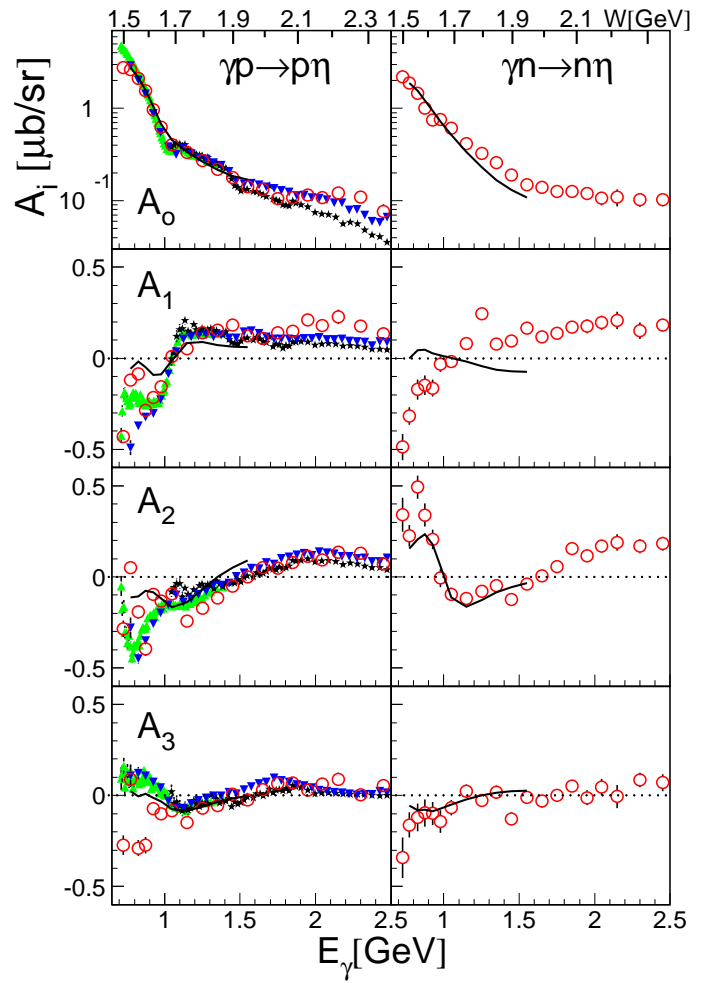}  
}
\caption{Coefficients of the Legendre series of the fits in 
Figs. \ref{fig:p_diff}~\ref{fig:ndiff},
Left hand side: quasi-free proton. 
Open (red) circles: present quasi-free data,
(blue) downward triangles: Bonn ELSA data \cite{Crede_09},
(black) stars: CLAS data \cite{Williams_09},
(green) upward triangles: Mainz MAMI data \cite{McNicoll_10}.
Right hand side: quasi-free neutron (from neutron coincidence $\sigma_n$).
Open (red) circles: present quasi-free data,
solid lines: Eta-MAID \cite{Chiang_02}.
Note the logarithmic scale for $A_0$. 
}
\label{fig:coeff}       % Give a unique label
\end{figure}

Around incident photon 
energies of 1 GeV, corresponding to $W\approx$~1.7 GeV a bump-like
structure is visible in the neutron cross 
section, which is not seen for the
proton. In fact, it is even more pronounced in the ratio of neutron and proton
data, which shows a sharp rise around $E_{\gamma}$=1~GeV.
Before we discuss this structure in detail, we compare the angular
distributions of the two reactions, which are summarized in Fig. \ref{fig:ndiff}.
They have been fitted with the Legendre series from Eq. \ref{eq:legendre}
and the coefficients are shown in Fig. \ref{fig:coeff}. 

The coefficients of the
quasi-free proton data are in good agreement with the free proton data as
expected from Fig. \ref{fig:p_diff}. The largest deviations occur for a few 
values of the $A_3$ coefficient close to the threshold, this might be due to 
uncorrected Fermi motion effects.

The neutron results obtained from the analysis with coincident recoil neutrons
($\sigma_n$) and from the difference of inclusive data and data with coincident
protons $\sigma_{n}'=\sigma_{np}-\sigma_p$ (filled and open circles in 
Fig. \ref{fig:ndiff}) are also in good overall agreement, some discrepancies occur
for the extreme angles, in particular in forward direction. 
Here, one should note that the detection efficiency for the reaction with coincident 
recoil protons almost vanishes for the extreme forward angles of the $\eta$-meson
(see Fig. \ref{fig:deffi}), so that $\sigma_{n}'$ is less well defined in this regime. 

For the comparison of proton and neutron cross sections we discuss three 
different energy ranges. At high incident photon energies above 1.5 GeV,
the absolute magnitude of the cross sections as well as the strongly forward 
peaked shape of the angular distributions are almost identical. This is the 
energy region, were previous model analyses of free proton data 
(see e.g. \cite{Crede_05}) have identified dominant contributions from 
$t$-channel background terms. These contributions seem to be similar for protons 
and neutrons, which is not unexpected when considering isospin invariance. 
Agreement with the MAID model at higher incident photon energies is not good,
the observed forward peaking of the cross section is not reproduced. The analysis 
of the previous ELSA proton data \cite{Crede_05,Bartholomy_07,Crede_09} in the 
framework of the Bonn-Gatchina model had found a strong contribution of a 
D$_{15}$(2070) state to $\eta$-production at large incident photon energies. 
This state is not visible as a bump in the total cross section, not even like 
the small indication of the P-resonances at $W\approx$1.7 GeV). It was extracted 
from the analysis of the angular distributions. The present neutron/proton ratio 
of the total cross section might show some structure in this energy region 
(cf. Fig. \ref{fig:ntot}, bottom part), however at the very limit of statistical
significance.

The total cross section for quasi-free $\eta$-production off the proton and the 
neutron are quite different at low incident photon energies (below 900 MeV) in the 
range of the second resonance region. In this regime, the absolute magnitudes of the 
cross sections reflect the ratio of the helicity amplitudes of the S$_{11}$(1535) 
resonance (see Eqs. \ref{eq:rat},\ref{eq:helrat}). The shape of the angular 
distributions is dominated by the interference between the S$_{11}$(1535) and the 
D$_{13}$(1520) resonances, which involves a term proportional to the A$_{2}$ 
coefficient of the Legendre series. The electromagnetic helicity-1/2 couplings 
$A_{1/2}$ of the D$_{13}$ have identical signs for proton and neutron, while the 
S$_{11}$ couplings have different signs.

\clearpage

\noindent{Therefore,} the interference term changes 
sign from proton to neutron, giving rise to negative A$_{2}$ coefficients for the 
proton and positive ones for the neutron \cite{Weiss_03}. This effect is also 
reflected in the model calculations \cite{Chiang_02,Shklyar_07,Shyam_08}. 

In the most interesting range around 1 GeV incident photon energy, where the
peak-like structure appears in the total neutron cross section, proton and 
neutron angular distributions are not very different (apart from the absolute 
scale). In this region, the S$_{11}$ - D$_{13}$ interference is not  visible any 
more ($A_{2}$ for proton and neutron is small and negative) and the $A_{1}$ 
coefficient shows a zero crossing with very steep rise. It was already 
discussed in \cite{Denizli_07} that the simplest explanation for a rapidly 
varying $A_{1}$ coefficient is an interference between $S$ and $P$ waves. 
This explanation seems to be natural since it is well known that in this energy
region the tail of the S$_{11}$(1535) resonance, the S$_{11}$(1650) resonance, 
and the P$_{11}$(1710) and/or P$_{13}$(1730) resonances contribute. 
The model analyses still disagree in the relative importance of the two $P$-wave 
states. The $\eta$-MAID model \cite{Chiang_02} finds a dominant contribution from 
the P$_{11}$, while the Bonn-Gatchina analysis \cite{Anisovich_05} prefers the
P$_{13}$ state. Very preliminary results from a measurement of the
$\vec{\gamma}\vec{p}\rightarrow p\eta$ reaction (circularly polarized beam, 
longitudinally polarized target) at ELSA \cite{Elsner_10} indicate 
a dominant helicity-1/2 component in this energy region, which is more in 
line with the MAID analysis favoring the P$_{11}$ contribution. 
If one assumes only S$_{11}$ and P$_{11}$ states ($E_{0^+}$ and $M_{1^-}$ 
multipoles), the $A_{1}$ coefficient would simply be proportional to
Re$(E_{0^+}^{\star}M_{1^-})$ and thus the fast zero crossing would 
imply a rapid phase change between these multipoles indicating, that one
is going through resonance around 1 GeV.

It is, however, not yet understood, whether the structure observed in the 
total neutron cross section at the same incident photon energy 
is somehow related to contributions from these resonances. Structures in the 
total cross section can obviously not arise from interferences between 
different partial waves. Therefore, different scenarios involving contributions 
from specific resonances as well as interference patterns in the same partial 
wave have been discussed in the literature. Fix, Tiator, and Polyakov
\cite{Fix_07} have investigated whether the data could be consistent with
the excitation of a narrow P$_{11}$-state. This work was motivated by the idea 
that the P$_{11}$-state of the proposed anti-decuplet of pentaquark states
should be relatively narrow (width on the order of 10 MeV), have a strong
electromagnetic coupling to the neutron, and a large $\eta N$ decay branching 
ratio \cite{Polyakov_03,Arndt_04}. They used two different versions of
the $\eta$-MAID model, the standard version \cite{Chiang_02} and the reggeized
version \cite{Chiang_03} as basis of their fits. The standard version, 
including a large contribution from the D$_{15}$(1675) resonance, reproduces
fairly well the experimental ratio of neutron and proton cross sections,
although it does not show the structure observed in the neutron data around 
photon energies of 900 MeV (see also discussion in the next section). 
The reggeized
version with a much smaller contribution of the D$_{15}$ reproduces the data 
only when an additional narrow resonance is introduced (taken as P$_{11}$).
Since the data are smeared out by Fermi motion, the width of this additional 
state is uncertain and could be as narrow as 10 MeV (roughly 40 MeV as upper
limit) \cite{Fix_07}. Similarly, an analysis performed in the framework of
the Bonn-Gatchina model (BoGa) \cite{Anisovich_09} can reproduce the neutron 
data reasonably well with three completely different scenarios, by either adding 
a `conventionally' broad P$_{11}$ resonance, a very narrow P$_{11}$ state, or
even by a careful adjustment of the interference pattern for the $S$-wave
amplitudes. Shklyar, Lenske, and Mosel \cite{Shklyar_07} find 
solutions in the framework of the Giessen coupled channel model with bump-like
structures in the neutron excitation function around $E_{\gamma}\approx$ 1~GeV
just from coupled channel effects in the S$_{11}$ - P$_{11}$ sector, without 
introducing any additional resonance. 
A similar result from a coupled-channels K-matrix approach was presented by 
Shyam and Scholten \cite{Shyam_08} who report a bump-like structure arising
from superpositions and interferences of contributions from the S$_{11}$(1535), 
S$_{11}$(1650), P$_{11}$(1710), and P$_{13}$(1720) states. However, this
structure appears broader than our experimental results discussed below. 
Finally, D\"oring and Nakayama
\cite{Doering_10}, using an $S$-wave coupled channel model, find a `dip-bump'
structure in the neutron cross section related to the opening strangeness 
thresholds of $K\Lambda$ and $K\Sigma$ photoproduction around 900 MeV and 
1050 MeV. Such unitary cusps are for example well-known for pion production 
reactions. The cusp structure in $\pi^0$ and $\pi^+$ photoproduction off the 
proton at the $\eta$-production threshold was discussed in detail in 
\cite{Althoff_79}, and the cusp structure in $\gamma p\rightarrow p\pi^0$ at the 
$\gamma p\rightarrow n\pi^+$ threshold was analyzed in \cite{Bernstein_97}.     

From the experimental side obviously two pieces of information
are missing to distinguish between these different scenarios. In the absence 
of any results for polarization observables, it is impossible to isolate the
responsible partial wave(s). Measurements of polarization observables like
the helicity asymmetry $E$ (longitudinally polarized target, circularly 
polarized beam), the target asymmetry $T$ (transversely polarized target), and 
the asymmetry $F$ (transversely polarized target, circularly polarized beam)
have been initiated at the Bonn and Mainz accelerators, for $\eta$ photoproduction
off the free proton also at the CLAS experiment at Jlab. Furthermore, the influence 
of the momentum distribution of the 
bound neutron obscures the intrinsic shape of the bump structure for
the free neutron. To overcome this difficulty, we present in the following
section a new analysis of the present data, based not on the incident photon
energy but on the invariant mass of the $\eta - N$ final state, which is not
affected by Fermi motion.   

\subsection{The $\eta$ - nucleon invariant mass distributions}  

In the previous sections we discussed the cross sections as function of the
incident photon energy measured with the tagging spectrometer. Due to the 
momentum distribution of the bound nucleons, each value of incident photon
energy corresponds to a broad distribution of invariant masses $W$ of the 
$\eta$ - participant-nucleon pairs, giving rise to the Fermi smearing of all
narrow structures. However, in principle we can directly extract $W$ from
the four-momenta of the $\eta$ and the participant nucleon. We have already 
shown in \cite{Jaegle_08} that the bump around 1.7 GeV in $W$ becomes then much
more narrow. In that analysis only data with $\eta$-mesons emitted at backward
angles were included, since then the kinetic energy of the neutron detected in
the TAPS detector can be determined from a time-of-flight measurement.
However, this analysis can be extended to the full data set, using the
kinematical overdetermination of the data.  

\begin{figure}[th]
\resizebox{0.5\textwidth}{!}{%
% normal version
  \includegraphics{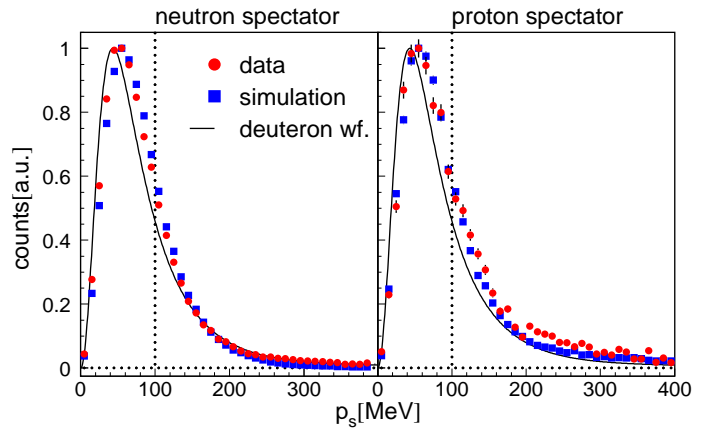}  
}
\caption{Momentum distributions of spectator nucleons.
(Red) dots: reconstructed from data, (black) lines: expected from deuteron
wave function \cite{Lacombe_81}, (blue) squares: Monte Carlo simulation
including detector response.
Left hand side: neutron spectator (i.e. recoil proton detected),
right hand side: proton spectator (i.e. recoil neutron detected). 
}
\label{fig:nferm}       % Give a unique label
\end{figure}

All kinematic variables (incident photon beam energy, target deuteron at rest)
of the initial state are known. For the final state
the four-momentum of the $\eta$-meson, the mass of participant and spectator
nucleon and the recoil direction of the participant nucleon 
(polar angle $\Theta$ and azimuthal angle $\Phi$) are known. Missing is only 
the three-momentum of the spectator- and the kinetic energy of the participant
nucleon. But these four variables can be extracted using energy and momentum
conservation which provide four equations. Monte Carlo simulations using the
GEANT package have shown that with this reconstruction a typical experimental 
resolution of FWHM$\approx$25 MeV for $W$ is achieved. 

As result of such an analysis Fig. \ref{fig:nferm} shows the distribution
of the momenta of the spectator nucleons, constructed event-by-event from
the reaction kinematics. In plane wave approximation with negligible FSI effects, 
these momenta must reflect the momentum distribution of the bound nucleons.
As demonstrated in the figure, this is quite well fulfilled. Data generated
with a participant - spectator Monte Carlo simulation using the deuteron wave
function as input and including the response of the detector reproduce the 
measured distributions. For further analysis one can in principle cut away 
events with large spectator momenta, which are not close to quasi-free 
kinematics. However, as it turned out 
(see Figs. \ref{fig:w_ratio},~\ref{fig:w_ratio2}), such a cut does not have much 
impact (in particular not on the angular distributions) apart from reducing 
counting statistics, so that it was not used for the differential cross sections.  

For the absolute normalization of the total cross section as function of $W$,
the photon flux $dN_{\gamma}/dE_{\gamma}$ measured as function of incident 
photon energy $E_{\gamma}$, was folded with the nucleon momentum distribution
to obtain the flux $dN_{\gamma}/dW$ in dependence of the final state invariant
mass. 
The results for the total quasi-free cross sections of the proton and
the neutron with and without cut on spectator momentum are summarized in Figs.
\ref{fig:w_ratio},~\ref{fig:w_ratio2}.
For both analyses good agreement between the quasi-free proton data and the 
world data set of free proton data is found. The neutron data show a 
pronounced, narrow peak around $W\approx$ 1.7 GeV. The position of this peak
coincides with a dip in the proton excitation function. 

In the following we discuss in more detail first the range of the S$_{11}$(1535)
resonance and then the narrow structure in the neutron excitation function.
 
\subsubsection{The region of the S$_{11}(1535)$ resonance peak}

The region of the S$_{11}$(1535) is interesting for two reasons. First since 
this is a well studied state, it can serve as a test case for the extraction
of resonance parameters from quasi-free data with the above discussed
kinematical reconstruction. Furthermore, the results contribute to the discussion
up to which energy range the observed cross section for $\eta$-photoproduction
is dominated by the S$_{11}$(1535). 

\begin{figure}[th]
\resizebox{0.5\textwidth}{!}{%
% normal version
  \includegraphics{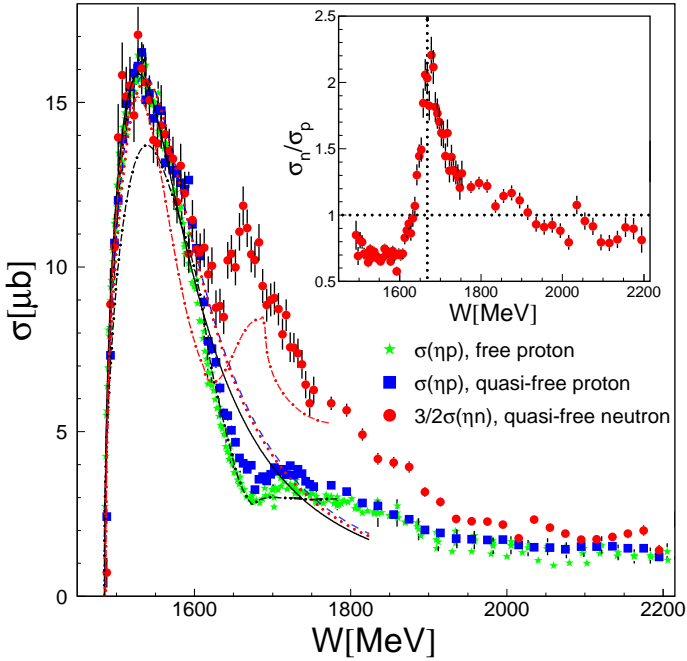}  
}
\caption{Total cross sections as function of final state invariant mass $W$
without cut on spectator momentum. (Red) dots: quasi-free neutron,
(blue) squares: quasi-free proton, (green) stars: free proton data.
Curves: fitted (up to $W$ = 1600 MeV) S$_{11}$(1535) line shapes. 
(Black) solid: free proton, (blue) dashed: quasi-free proton, (red) dotted:
quasi-free neutron. Dash-dotted curves: model results from \cite{Doering_10}.
Insert: ratio of quasi-free neutron - proton data. 
}
\label{fig:w_ratio}       % Give a unique label
\end{figure}

In the region of the S$_{11}$(1535) peak all data have been fitted with a 
parameterization of this resonance as Breit-Wigner curve with energy dependent 
width \cite{Krusche_03}:
\begin{equation}
\sigma(W)=\frac{q_{\eta}^{\star}}{k^{\star}}
\cdot\frac{k^{\star}_{R}}{q_{\eta R}^{\star}}
\cdot\frac{2m_{N}\cdot W_{R}\cdot b_{\eta}\cdot (A_{1/2}^{N})^2\cdot\Gamma_{R}}
{(W_{R}^2-W^2)^2+W_{R}^2\Gamma_R^2 x^2}
\end{equation} 
with
\begin{equation}
x = b_{\eta}\cdot \frac{q_{\eta}^{\star}}{q_{\eta R}^{\star}}
  + b_{\pi} \cdot \frac{q_{\pi}^{\star}}{q_{\pi R}^{\star}} + b_{\pi\pi}
\end{equation}
where $W_{R}$, $\Gamma_{R}$ are resonance position and width, 
$k_{R}^{\star}$, $q_{\eta R}^{\star}$, $q_{\pi R}^{\star}$ are incident 
photon momentum, $\eta$-momentum and $\pi$-momentum at resonance position
in the photon - nucleon-at-rest cm-frame, $A_{1/2}^{N}$
is the electromagnetic helicity-1/2 coupling, $b_{\eta}$=0.5, $b_{\pi}$=0.4, 
and $b_{\pi\pi}$=0.1 have been used as branching ratios for the $N\eta$,
$N\pi$, and $N\pi\pi$ decays of the resonance, and $m_N$ is the nucleon mass.  

\begin{table}[h]
\begin{center}
\caption{Result of Breit-Wigner fits. $N\eta$ branching ratio of S$_{11}$
is assumed as $b_{\eta}$=0.5. Upper part of table: Comparison of fits of
S$_{11}$(1535) resonance for free proton, quasi-free proton, and quasi-free
neutron data to PDG estimates \cite{PDG} and BoGa model fit \cite{Anisovich_09}. 
$\dag$: Breit-Wigner mass, in brackets pole position; $\ddag$: only magnitudes,
no signs; $\S$: only pole positions given, no Breit-Wigner mass.
Bottom part: fit of neutron data with S$_{11}$. 
resonance and two further Breit-Wigner curves. All uncertainties of fit
parameters are statistical only.
%$A_{1/2}\cdot\sqrt{b_{\eta}}$, 39$\pm$6, 12$\pm$3
} 
\label{tab:bwfit}       % Give a unique label
% For LaTeX tables use
\begin{tabular}{|c|c|c|c|}
\hline\noalign{\smallskip}
S$_{11}$(1535) & $W^{\dag}$ [MeV] & $\Gamma^{\dag}$ [MeV] & $A_{1/2}^{\ddag}$ \\
& & & [10$^{-3}$GeV$^{-1/2}$] \\
\hline
PDG & 1535$\pm$10 & 150$\pm$25 & $A_{1/2}^p$: 90$\pm$30 \\
 & (1510$\pm$10) & (170$\pm$80) & $A_{1/2}^n$: 46$\pm$27 \\
 \hline
BoGa$^{\S}$ & - & - &  $A_{1/2}^p$: 90$\pm$25 \\ 
& (1505$\pm$20) & (145$\pm$25) & $A_{1/2}^n$: 80$\pm$20 \\
\hline
$\gamma p\rightarrow p\eta$ & 1536$\pm$1 & 170$\pm$2 & 106$\pm$1 \\
\hline
$\gamma d\rightarrow (n)p\eta$ & 1544$\pm$2 & 181$\pm$13 & 109$\pm$3 \\
\hline
$\gamma d\rightarrow (p)n\eta$  & 1546$\pm$3 & 176$\pm$20 & 90$\pm$4  \\
\hline
\hline
$\gamma d\rightarrow (p)n\eta$  &  &  &  \\
\hline
S$_{11}$(1535)  & 1535$\pm$4 & 166$\pm$23 & 88$\pm$6 \\
\hline
`broad BW' & 1701$\pm$15 & 180$\pm$35 & - \\
\hline
`narrow BW' & 1663$\pm$3 & 25$\pm$12 & -  \\
\noalign{\smallskip}\hline
\end{tabular}
\end{center}
\end{table}

The data have been fitted up to $W\approx$1.6 GeV, where the proton and neutron
cross sections start to deviate and obviously the line-shape can no longer be
dominated by the S$_{11}$-resonance. The fit results are shown in 
Fig. \ref{fig:w_ratio} and the parameters are summarized in the upper part of 
table \ref{tab:bwfit}. The results demonstrate the following. Breit-Wigner mass 
and width of the resonance extracted from the quasi-free proton and neutron data
are in excellent agreement. The agreement with the free proton data is good, but
not within statistical uncertainties. This had to be expected due to the finite
$W$ resolution of the quasi-free data, which tends to increase the width and
to shift the resonance position slightly upward. These parameters are also in
good agreement with the values given by the particle data group \cite{PDG}
and the Bonn-Gatchina analysis \cite{Anisovich_09} of the present data. Note 
that for the BoGa analysis not Breit-Wigner masses but pole positions are given,
which agree with the PDG parameters. The almost perfect agreement of the shape
of the S$_{11}$-peaks for the proton and neutron data in this range is also
further evidence that this shape is alone dominated by the S$_{11}$(1535).
Effects e.g. from the destructive interference of the two
S$_{11}$-resonances, which are important at higher incident photon energies,
should be different for proton and neutron since the ratios of the electromagnetic
couplings of these two resonances are quite different for protons and neutrons.  

The electromagnetic helicity couplings
$A_{1/2}^{N}$ found from the fits agree for free and quasi-free proton data.
The proton couplings are slightly higher than the PDG value and the BoGa result.
This could be a systematic effect, since all non-S$_{11}$(1535) contributions are
neglected in the fit. The only large discrepancy arises for the neutron helicity
coupling between the present and BoGa results on one hand side and the PDG value
on the other side, which is much lower. Here, one should note that as already
discussed in \cite{Krusche_03} there is a systematic discrepancy between the
helicity couplings of the S$_{11}$(1535) extracted from pion photoproduction 
versus those from $\eta$-photoproduction. The latter ones are significantly
larger. The S$_{11}$(1535) dominates $\eta$ production but contributes
only weakly to pion production which is dominated in this energy range by the
D$_{13}$(1520). Therefore, $\eta$-production is the better suited channel for the
study of the S$_{11}$(1535) properties. In the meantime the PDG proton coupling 
became dominated by the larger values from $\eta$ production, but the neutron
coupling is still dominated by the small values from pion production.
The resulting PDG neutron/proton ratio of the helicity couplings would correspond 
to a cross section ratio for $\eta$ production in the S$_{11}$ maximum of 0.26, 
which is unrealistic. The BoGa analysis finds a ratio of 0.79 and the simple 
BW-fits a ratio of 0.68.

\subsubsection{The region of the narrow peak in the $\gamma n\rightarrow n\eta$
reaction}

In order to estimate the width of the narrow structure observed in the neutron
data, the excitation function has been fitted up to $W\approx$ 1.8 GeV with a
purely phenomenological fit function. It is composed of the Breit-Wigner curve 
with energy dependent width for the S$_{11}$(1535) resonance and two further 
simple Breit-Wigner curves with constant width ($x\equiv$1). The curves are
compared to the data in Fig.~\ref{fig:w_ratio2} and the fit parameters are listed 
in the bottom part of Tab. \ref{tab:bwfit}. The parameters obtained for the
S$_{11}$ are consistent with the results discussed above. The broad BW-curve
located at $W\approx$ 1.7 GeV just serves for the
effective parameterization of the excitation function. 
It subsumes contributions from all normally broad 
resonances in this energy region (such as P$_{11}$(1710), P$_{13}$(1720), 
D$_{15}${1650),...) as well as background components.
The narrow Breit-Wigner curve at $W\approx$1.66 GeV has a FWHM of only 
(25$\pm$12) MeV, on the same order as the experimental resolution of 25 MeV (FWHM). 
This width is somewhat dependent on the chosen parameterization, but also trials 
with different background shapes, e.g. of polynomial type, which result in a 
poorer fit quality, indicate a width below the 50 MeV level. 

\begin{figure}[th]
\resizebox{0.5\textwidth}{!}{%
% normal version
  \includegraphics{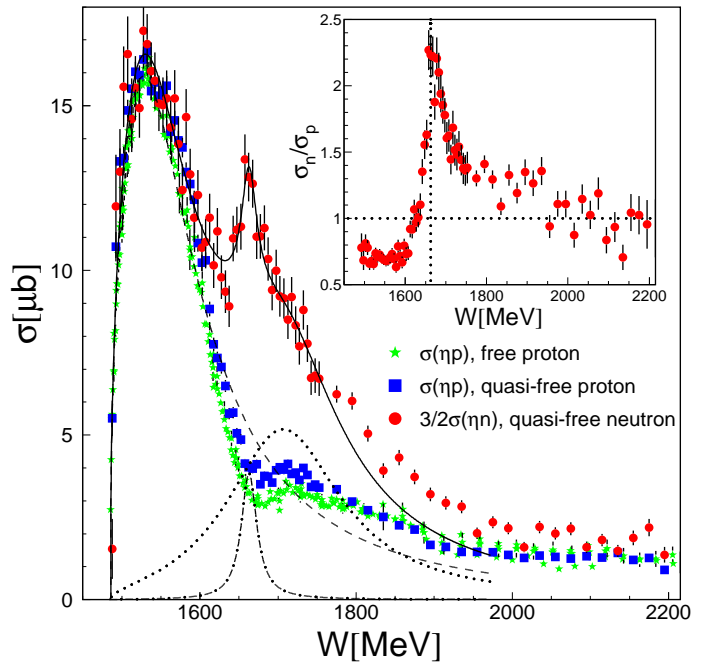}  
}
\caption{Total cross sections as function of final state invariant mass $W$
for spectator momenta $p_S<$ 100 MeV. Notation as in Fig. \ref{fig:w_ratio}.
All curves for neutron data; dashed: fitted S$_{11}$ line shape, dotted:
broad Breit-Wigner resonance, dash-dotted: narrow Breit-Wigner, solid:
sum of all. 
}
\label{fig:w_ratio2}       % Give a unique label
\end{figure}

Since so far there is no information about
the quantum numbers of this structure, in fact it is not even clear, whether it
corresponds to a nucleon resonance, parameters like electromagnetic couplings 
cannot be given. However, if we treat the structure as a narrow S$_{11}$ 
resonance the normalization of the fit corresponds to 
$A_{1/2}\cdot\sqrt{b_{\eta}}\approx 12\times 10^{-3}$GeV$^{-1/2}$.

The results for the angular dependence of the excitation functions are 
summarized in Figs. \ref{fig:dexi},~\ref{fig:w_coeff}. 
Due to statistical limitations in the extraction process of the cross sections
depending on the final state $W$, the angular distributions are only coarsely binned.
Fig. \ref{fig:dexi} shows in the upper part excitation functions in dependence 
on $W$ for four different bins of cm polar angles as well as the neutron/proton 
ratios. The bottom part shows the corresponding angular distributions for 
different bins of $W$. Finally, Fig. \ref{fig:w_coeff} summarizes coefficients 
of the Legendre series of Eq. \ref{eq:legendre}, fitted to the angular 
distributions. The results are compared to free proton data and model
calculations. The comparison to free proton data from the recent most precise 
measurement at MAMI \cite{McNicoll_10} demonstrates impressively how well the 
elementary reaction on the free proton can be approximated by quasi-free data
with $W$ reconstructed from the $p\eta$ final state kinematics. 
Significant deviations occur only very close to the production threshold,
where the effects from Fermi motion are most pronounced. 
Comparing proton and neutron data, in particular the excitation functions for
forward angles, it is even more apparent than in the total cross section that 
the narrow structure observed in the neutron excitation function is accompanied 
by a pronounced dip in the proton data at the same position and of comparable 
width. It seems to be highly unlikely that these two structures are unrelated.
This might indicate that some interference with a sign change between proton
and neutron is involved.

\begin{figure*}[th]
\center{\resizebox{0.95\textwidth}{!}{%
  \includegraphics{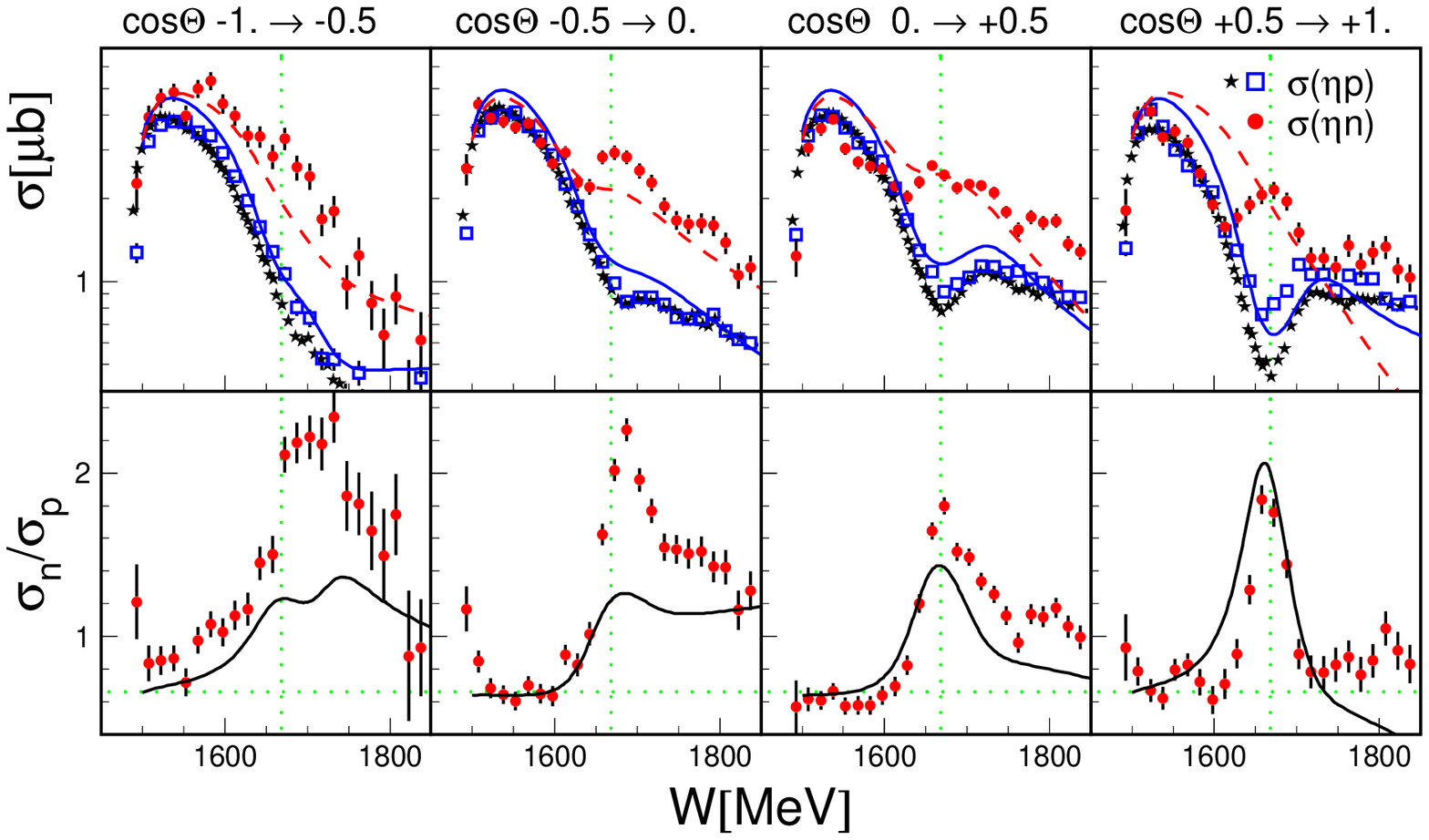}
}}

\vspace*{0.5cm}
\center{\resizebox{0.95\textwidth}{!}{%    
  \includegraphics{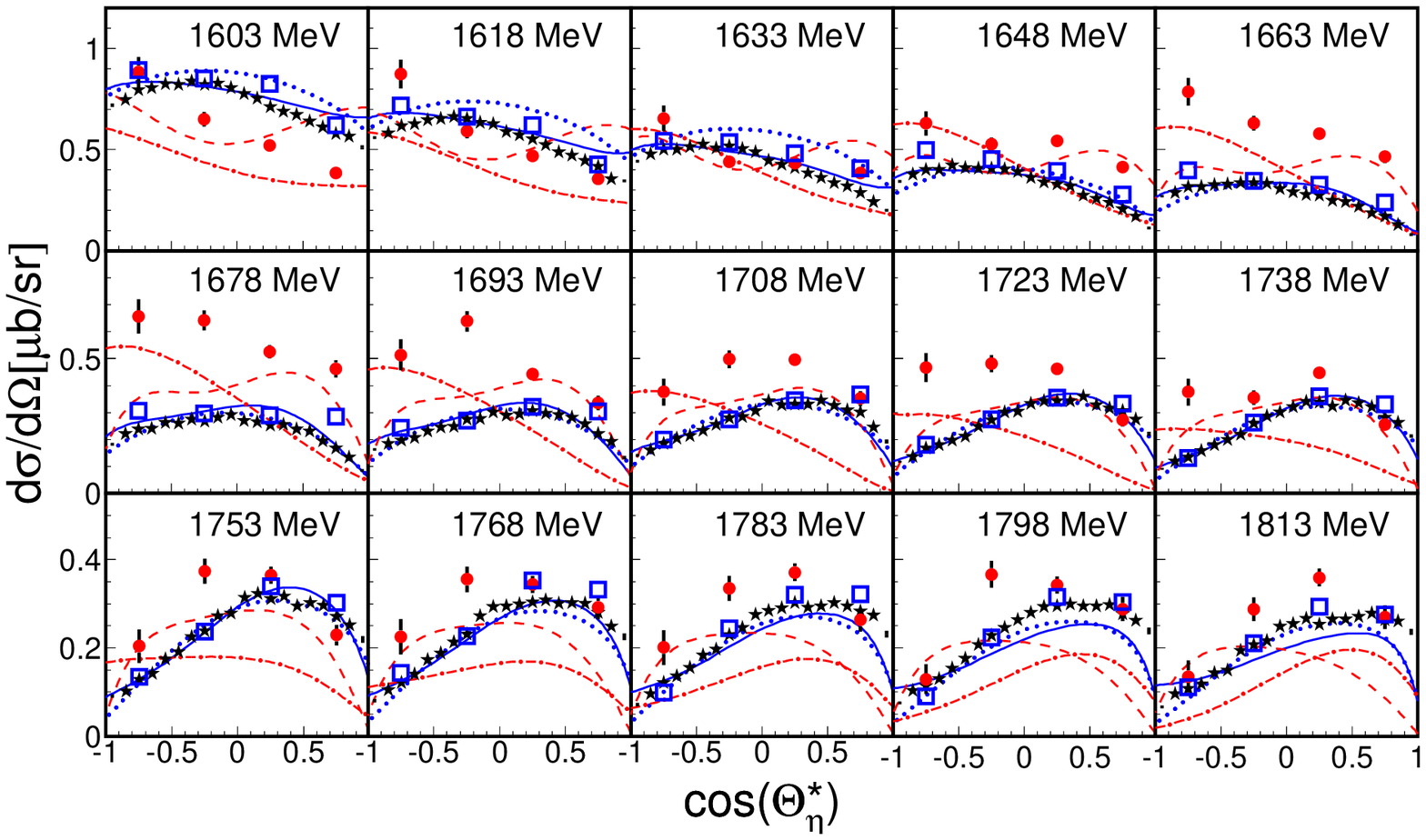}    
}}
\caption{Top, first row: excitation functions for different bins of $\eta$ cm 
polar angle.
(Blue) open squares: quasi-free proton data, (black) stars: free proton data 
from \cite{McNicoll_10}, (red) dots: quasi-free neutron data scaled up by 3/2. 
(Blue) solid lines: $\eta$-MAID \cite{Chiang_02} for the proton target, 
(red) dashed lines: $\eta$-MAID for the neutron target. 
Second row: ratio of neutron and proton cross section for data and $\eta$-MAID.
Vertical dotted lines: position of narrow peak in neutron data,
horizontal dotted lines: $\sigma_n/\sigma_p$=2/3.
Bottom part: angular distributions. Same notation as top part (but neutron data 
not scaled); additional
curves (blue) dotted: proton model from Shklyar et. al. \cite{Shklyar_07},
(red) dash-dotted: neutron model from same reference.
}
\label{fig:dexi}       % Give a unique label
\end{figure*}
\clearpage

\begin{figure}[th]
\resizebox{0.5\textwidth}{!}{%
% normal version
  \includegraphics{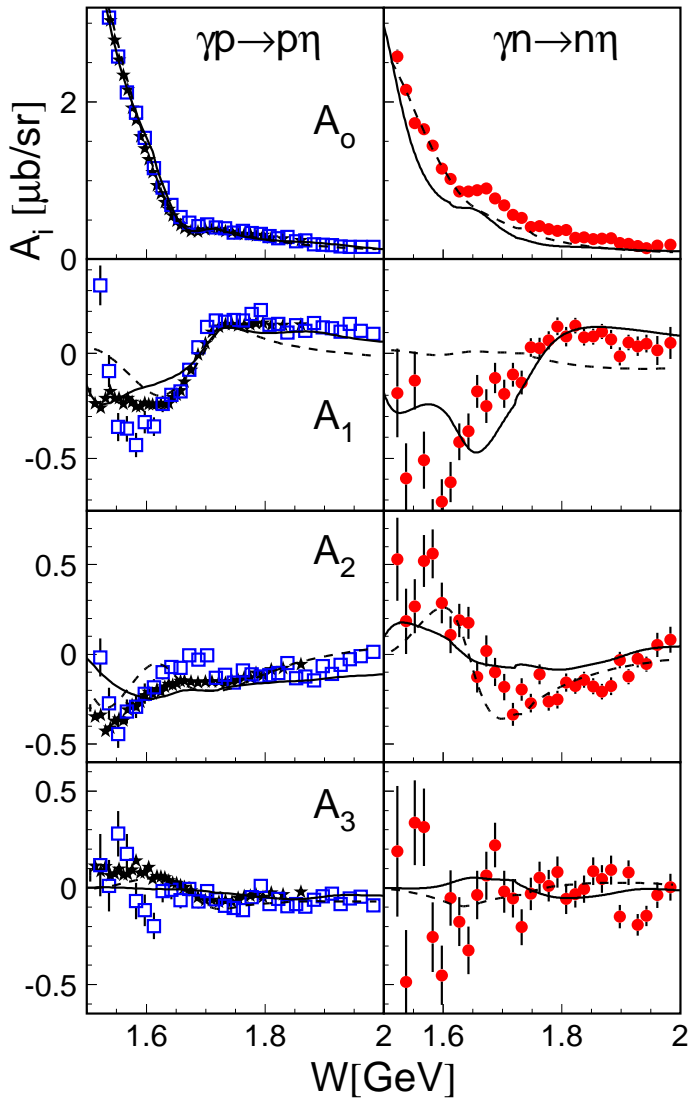}  
}
\caption{Fitted coefficients of the Legendre series as function of final state
invariant mass $W$. Left hand side: 
(black) stars free proton, (blue) squares quasi-free proton. Right hand side:
(red) dots quasi-free neutron. Curves: dashed standard MAID model
\cite{Chiang_02}, solid: Shklyar et al. \cite{Shklyar_07}.
}
\label{fig:w_coeff}       % Give a unique label
\end{figure}

\noindent{However,} since the neutron peak and the proton dip are
also visible in the total cross section, at least part of this interference
must be in the same partial wave. A dip-like structure has also been observed
for comparable values of $W$ in the $\pi^- p\rightarrow \eta n$ reaction, 
although at much lower statistical significance. As a possible explanation,
similar to one version of the BoGa-model \cite{Anisovich_09}, the
interference between the two S$_{11}$ resonances was discussed in a $K$-matrix
approach \cite{Gridnev_99}.
  
The comparison of the neutron data to model predictions leads to the
following conclusions. Models which try to explain the structure observed 
in the Fermi smeared excitation function of the neutron data by one
conventionally broad nucleon resonance like one of the scenarios in
\cite{Anisovich_09} are ruled out by the narrow width 
(see Fig.~\ref{fig:w_ratio2}) on the order of 25 MeV. 

The coupled channel
approach of D\"oring and Nakayama \cite{Doering_10} shown in Fig.
\ref{fig:w_ratio} in fact produces a structure of similar width close to the
observed position, although the exact shape is somewhat different. Since this
model includes only s-wave contributions it cannot predict realistic shapes of
angular distributions.
Nevertheless, the fitted Legendre coefficients (see Fig. \ref{fig:w_coeff})
do at least not contradict the assumption that only s-waves are important. 
The
peak-like structure is clearly seen only in $A_{0}$, which is proportional to
the total cross section. From the higher coefficients only $A_{3}$ might 
show a little structure close by, albeit not statistically significant.

A direct comparison of the predictions of the MAID model \cite{Chiang_02} and 
the model from Shklyar et al. \cite{Shklyar_07} to the angular distributions 
(see bottom part of Fig.~\ref{fig:dexi}) shows that both models disagree with 
the neutron data throughout most of the energy range, in particular around
the peak structure at $W\approx$ 1.67 GeV. Agreement with the proton data is
of course much better since free proton data have been used to fix the
parameters of both models. 

A more detailed comparison reveals some interesting
features. The MAID model reproduces reasonably well the prominent structure 
in the ratio of neutron and proton cross section in the forward angular range
(upper part of Fig. \ref{fig:dexi}). However, even there it does not show any
indication of the peak-like structure in the excitation function of the
$\gamma n\rightarrow n\eta$ reaction. The peak in the ratio stems alone from the 
dip in the proton cross section. Since
it is unlikely that these two structures are unrelated, this casts
also some doubts whether the `dip' structure in the proton cross section
was correctly interpreted in this model.
The comparison certainly rules out 
that the neutron structure can be entirely explained by a strong contribution 
of the D$_{15}$(1675) resonance. A comparison of the fitted Legendre
coefficients shows that the MAID model does not reproduce the peak in the
$A_{0}$ coefficients and also fails completely for $A_{1}$, indicating that
the strong $S-P$ interference is not reflected in the model. On the other hand,
the good agreement for the $A_{2}$ coefficient means that the $S-D$ 
interference term (in particular the S$_{11}$(1535) - D$_{13}$(1520) 
interference) is well understood. The Shklyar model \cite{Shklyar_07} shows at
least some indication for a peak in $A_{0}$ at the right position and is in much
better agreement with $A_{1}$, although here it predicts a dip structure around
$W\approx$1.67 GeV, which is not in the data. Agreement with $A_{2}$ is not
as good as for MAID. It is evident that the comparison of data and
models does not allow a final conclusion about the nature of the structure
in the neutron excitation function.  

\section{Summary and Conclusions}

Precise cross section data have been measured for quasi-free photoproduction of
$\eta$-mesons off protons and neutrons bound in the deuteron. Due to the 
combined analysis of events in coincidence with recoil protons, recoil 
neutrons, and of the inclusive reaction systematic uncertainties related to the 
detection of the recoil nucleons could be reliably controlled. 
The results confirm earlier measurements in the region of the S$_{11}$(1535) 
resonance and reveal an unexpected structure in the total cross section of
the $\gamma n\rightarrow n\eta$ reaction around incident photon energies of 1~GeV. 

The results of an analysis based on the invariant mass of the $\eta$-proton 
pairs from quasi-free photoproduction off the proton are in excellent agreement
with free proton data. This demonstrates that no nuclear effects beyond Fermi
motion are involved and that these effects can be reliably removed by this
method. An identical analysis of the quasi-free neutron data confirms a very
narrow peak in the neutron excitation function around $W\approx$1.67 GeV
with a width of only $\approx$25 MeV (FWHM), which seems to correspond to a dip-like
structure in the proton excitation function at the same energy. The nature of this 
structures is not yet understood.
Clearly ruled out are only single isolated 
resonances with conventional width like e.g. the D$_{15}$(1675) in the MAID model. 
Scenarios with a narrow resonance or different types of coupled channel effects
cannot yet be discriminated with the available data.

\section{Acknowledgments}
We wish to acknowledge the outstanding support of the accelerator group
and operators of ELSA. 
We gratefully acknowledge stimulating discussions with D. Glazier
concerning the analysis of the final state invariant mass and with
M. D\"oring, V. Shklyar, and L. Tiator about the interpretation of the results.
This work was supported by Schweizerischer Nationalfonds and
the Deutsche For\-schungs\-ge\-mein\-schaft (SFB/TR-16.)

\end{document}